\documentclass[1p]{elsarticle}

\journal{J. Logic. Algebr. Program}

\usepackage{tikz}
\usetikzlibrary{calc, positioning}
\usepackage{amsmath}

\usepackage{ifxetex}

\ifxetex
\usepackage{fontspec}
\usepackage{unicode-math}
\else
\usepackage{amssymb}
\usepackage[utf8]{inputenc}
\usepackage[T1]{fontenc}
\newcommand\coloneq{:=}
\fi
\usepackage[noliterate]{maude}
\usepackage[colorlinks]{hyperref}
\usepackage[capitalize,noabbrev]{cleveref}
\makeatletter
\let\c@author\relax
\makeatother

\newcommand\N{\ensuremath{\mathbb{N}}}

\newcommand\ao{\,\lower1pt\hbox{\normalfont @}\,}

\newcommand\action[1]{\emph{#1}}
\newcommand*\kywd[1]{\textsf{\bfseries #1}}
\newcommand*\skywd[1]{\textsf{\color{darkgray}\bfseries #1}}

\ifxetex
\newcommand\ctlAllw{\ensuremath{\mdlgwhtsquare\,}}
\else
\newcommand\ctlAllw{\ensuremath{\Box\,}}
\fi

\newcommand\refcd[1]{}

\makeatletter
\def\ps@pprintTitle{\let\@oddhead\@empty
     \let\@evenhead\@empty
     \def\@oddfoot
       {\hbox to \textwidth {\ifnopreprintline\relax\else
        \@myfooterfont \ifx\@elsarticlemyfooteralign\@elsarticlemyfooteraligncenter \hfil\@elsarticlemyfooter\hfil \else \ifx\@elsarticlemyfooteralign\@elsarticlemyfooteralignleft \@elsarticlemyfooter\hfill{}\else \ifx\@elsarticlemyfooteralign\@elsarticlemyfooteralignright {}\hfill\@elsarticlemyfooter \else \normalshape\hfill\begin{tikzpicture}
			\node at (0, 2.5em) {};
			\node[draw=black!70, fill=black!5, inner sep=5pt, text width=.855\linewidth]{
				Accepted authors' manuscript of the article published in \@journal\ 124 \\
				DOI: \href{https://doi.org/10.1016/j.jlamp.2021.100728}{10.1016/j.jlamp.2021.100728} \hfill License: CC-BY-NC-ND
			};
		\end{tikzpicture} \hfill\fi \fi \fi \fi }
       }\let\@evenfoot\@oddfoot}
\makeatother

\begin{document}

\begin{frontmatter}

\address[ucm]{Facultad de Informática, Universidad Complutense de Madrid, Spain}
\address[itc]{Instituto de Tecnología del Conocimiento, Universidad Complutense de Madrid, Spain}

\cortext[cor1]{Corresponding author}

\title{Metalevel transformation of strategies}
\author[ucm]{Rubén Rubio\corref{cor1}}
\ead{rubenrub@ucm.es}
\author[ucm,itc]{Narciso Martí-Oliet}
\ead{narciso@ucm.es}
\author[ucm]{Isabel Pita}
\ead{ipandreu@ucm.es}
\author[ucm]{Alberto Verdejo}
\ead{jalberto@ucm.es}

\begin{abstract}
	In the reflective Maude specification language, based on rewriting logic, a strategy language has been introduced to control rule rewriting while avoiding complex and verbose metalevel programs. However, just as multiple levels of reflection are required for some metaprogramming tasks, reflective manipulation and generation of strategies are convenient in multiple situations. Some examples of reflective strategy transformations are presented, which implement special forms of evaluation or extend the strategy language while preserving its advantages.
\end{abstract}


\end{frontmatter}

\section{Introduction}

	Reflection can be intuitively defined as the capacity of a system for reasoning about itself, by representing and manipulating its objects in its own language.
Classical examples of reflection can be seen in the coding of first-order arithmetic by Gödel and in universal Turing machines, but reflective metaprogramming features are also provided by many modern programming languages~\cite{metaprogrammingSurvey}. Rewriting logic~\cite{rewritingLogic} and its implementation Maude~\cite{maude} are reflective languages where important aspects of its own metatheory can be represented~\cite{reflection2007}. As a result, manipulating, transforming, and analyzing rewriting logic theories specified in Maude can be easily done within Maude.
Reflection has been extensively used throughout the history of Maude for specific metalinguistic applications, to extend and prototype new features of the language, and to design formal tools that reason about Maude programs. Significant examples are Full Maude~\cite[Part {II}]{maude} and the Maude Formal Environment~\cite{mfe}. The former is an extended Maude interpreter written in Maude itself, and the latter allows checking properties like confluence and termination on Maude specifications.

	Rewriting systems are executed by successive and independent rule applications where rule and position are chosen nondeterministically. However, it is sometimes convenient to restrict and control how rules are applied either for semantic or efficiency purposes. This can be expressed at a higher level and without modifying the original system by means of~\emph{rewriting strategies}~\cite{allthat,extstrat}. Classical examples are the different reduction strategies of the $\lambda$-calculus~\cite{barendregt} and those guiding deduction procedures and theorem provers~\cite{satstrats,lescanneOrme}. Moreover, strategies are a useful resource to write compositional rewriting specifications where the concerns of rules and their control are separate~\cite{lescanneOrme}, so that the same rules can yield different algorithms depending on how they are applied. Some languages of programmable strategies have been developed to express rewriting strategies in an executable form like ELAN~\cite{elan}, Stratego~\cite{stratego} for program transformation, TOM~\cite{tom}, $\rho$Log~\cite{rholog}, and Porgy~\cite{porgyJournal} for graph-rewriting. In Maude, the metaprogramming features have been traditionally used to program the control of rules. Since programming metalevel computations is hard for beginners and verbose, an object-level strategy language has been proposed, tested, and finally made available in Maude~3~\cite{maude}. Although the strategy language has been introduced to avoid the need for the metalevel, the language itself and its operations have been metarepresented, and users may still resort to the metalevel to analyze strategy specifications and construct strategies depending on metatheoretic information.
These transformed strategies can still be used at the object level thanks to the Maude support for interactive interfaces, and be analyzed using verification tools like the model checker for systems controlled by strategies~\cite{fscd,btimemc}.
	 
In this paper, we aim to show through three relatively simple examples the interesting applications and potential of the reflective manipulation and generation of strategies, and the resources offered by the Maude specification language to do so. Metaprogramming strategies is useful in multiple situations where they should be adapted to some input data specification or to the rewriting system being controlled itself. The first example in~\cref{sec:1stexample} should be understood as an introduction to the tools and the approach proposed in the paper, which can be applied to many other specific problems like this. In this case, given a Maude specification where operators are annotated with some restrictions, we generate a well-known normalization strategy in context-sensitive rewriting~\cite{contextRew} that can be applied to any term. The second example in~\cref{sec:langext} describes a general procedure to extend the Maude strategy language with new combinators without losing any of its advantages. A skeleton is provided in order to simplify the task of the extension developer and facilitate the interaction of its users. This is illustrated with two families of operators available in other strategy languages. Finally, the third example in~\cref{sec:multistrategies} presents a framework to specify compositional or agent-based strategy-controlled systems, whose control by a single strategy expression is cumbersome. Separate strategies are assigned to each agent or component, and a global strategy orchestrates their execution (concurrently, by turns, or as desired). Not only we think that these examples illustrate the possibilities of reflective transformations, but they are interesting by themselves. Although reflective languages featuring strategies are not common, this approach could be used in other tools apart from Maude, as it consists of applying the advantages of program transformation to this specific setting.

	\Cref{sec:prelim} begins by reviewing the basic of rewriting and Maude, before the aforementioned examples are introduced in~\cref{sec:1stexample,sec:langext,sec:multistrategies}. \Cref{sec:conclusions} presents a discussion on related work and the conclusions.
Maude~3 can be downloaded from \href{http://maude.cs.illinois.edu}{\ttfamily maude.cs.illinois.edu}, and its extension with the strategy-aware model checker is available at \href{http://maude.ucm.es/strategies}{\ttfamily maude.ucm.es/strategies}, as well as all the different examples appearing here.

\section{Rewriting logic and Maude} \label{sec:prelim}

	\emph{Rewriting logic}~\cite{rewritingLogic} was proposed as a unified model of concurrency extending membership equational logic with nondeterministic and possibly conditional rewriting rules. A rewrite theory $\mathcal R = (\Sigma, E, R)$ consists of a signature $\Sigma$ of order-sorted operators, i.e.\ their domains and values are typed with sorts and these sorts are related by a containment partial order relation, a set of equations $E$, and a set of rewriting rules $R$. Terms and rule applications are considered modulo equations and also structural axioms like commutativity, associativity, and identity that cannot be naively handled as regular equations due to their reversible nature. Maude~\cite{maude} is a specification language based on rewriting logic, where rewrite systems can be specified compositionally, executed, and analyzed. Specifications are written in a mathematical-like notation and organized in modules of different kinds: functional modules (\kywd{fmod}) represent equational theories with declarations of \kywd{sort}s, \kywd{subsort} relations, and \kywd{op}erators. Beside their signatures, operator declarations may include some attributes between brackets that specify the structural axioms and other features applied to them. Moreover, functional modules may include equations of the form:
\[ [\kywd{c}]\kywd{eq} \quad l \;\texttt{=}\; r \quad [\; \kywd{if}\; \bigwedge_i l_i \;\texttt=\; r_i \;\;\verb|/\|\;\; \bigwedge_i l'_i \;\texttt{:=}\; r'_i \;\;\verb|/\|\;\; \bigwedge_i t_i \;\texttt{:}\; s_i \;]\; \texttt. \]
Equations are applied as if they were oriented from left to right on any position where they match.\footnote{An unconditional oriented equation or rewriting rule $l \to r$ is applied to a term $t$ if there is a substitution $\sigma$ assigning terms to the variables of $l$ and a position $p$ in $t$ whose subterm $t|_p = \sigma(l)$. This subterm is then replaced by $\sigma(r)$. We say that $l$ matches $t_p$ and that $\sigma$ is the matching substitution.} Every variable in $r$ and the condition must occur in the left-hand side $l$ with some exceptions.
Conditions \texttt{$l_i$ = $r_i$} are satisfied when these terms, instantiated by the matching substitution, coincide modulo equations and axioms. The same semantics operates on the \texttt{$l'_i$ := $r'_i$} conditions, but the term $l'_i$ may contain free variables that are instantiated by matching and can be used in the following condition clauses. Sort-membership condition fragments $t_i : s_i$ hold when the instantiated term $t_i$ belongs to the sort $s_i$. For example, the following functional module specifies a list of letters:
\begin{lstlisting}
fmod LLIST-FM is
  protecting NAT .
  sorts Letter List .
  subsort Letter < List .

  ops a b c d e : -> Letter [ctor] .
  op nil : -> List [ctor] .
  op __ : List List -> List [ctor assoc comm id: nil] .

  var L : Letter . var LS : List .

  op length : List -> Nat .
  eq length(nil) = 0 .
  eq length(L LS) = 1 + length(LS) .
endfm
\end{lstlisting}
Underscores in operator names mark the holes where arguments are entered, although symbols in prefix notation may omit them. In \texttt{LLIST-FM}, \texttt{Letter} is made a subsort of \texttt{List}, the juxtaposition operator \lstinline{__} is declared associative, commutative, and having \texttt{nil} as identity, and the module \texttt{NAT} is imported. The Maude prelude provides some modules like \texttt{NAT} that specify integer and floating-point numbers, strings, lists, sets, etc. Importation can be done with the keywords \texttt{protecting}, \texttt{extending}, or \texttt{including} that declare whether the definitions of the imported module will be kept unchanged, extended, or modified arbitrarily, respectively. Functional specifications can be executed with the \texttt{reduce} command, which applies the equations exhaustively on the given term.
\begin{lstlisting}[language={}]
Maude> reduce length(a b c) .
rewrites: 7
result NzNat: 3
\end{lstlisting}

	System modules (\kywd{mod}) are rewrite theories with the addition of rules
\[ [\kywd{c}]\kywd{rl} \quad l \;\texttt{=>}\; r \quad [\; \kywd{if}\; C \;\verb|/\|\; \bigwedge_i\; l_i \;\texttt{=>}\; r_i \;]\; \texttt. \]
Conditional rules may include rewriting conditions in addition to those $C$ available for equations, where terms matching $r_i$ are searched by rewriting from $l_i$ with the rules of the module. Every possible match of the left-hand side and the condition yields a different application of the rule. Like equations, all variables in the right-hand side $r$ and the condition must occur in the left-hand side $l$, except those in the left-hand side of matching conditions and now also in the right-hand side $r_i$ of rewriting conditions, since these are assigned by matching. Continuing with the example, the following module \texttt{LLIST-M} extends the previous \texttt{LLIST-FM} with two rules.
\begin{lstlisting}
mod LLIST-M is
  including LLIST-FM .
  var LS : List . var L : Letter .

  rl [pop] : LS L => LS .
  rl [put] : LS   => LS L [nonexec] .
endm
\end{lstlisting}
Note that the \texttt{put} rule contains an unbounded variable \texttt{L} in its right-hand side. What would be a syntax error without the \texttt{nonexec} attribute, which excludes the rule from being applied, can be useful when combined with a strategy language able to instantiate this variable. Rules can be executed modulo the equations with the \texttt{rewrite} command in the Maude interpreter, which will repeatedly apply the \texttt{pop} rule in the previous module.
\begin{lstlisting}[language={}]
Maude> rewrite a b c .
rewrites: 3
result List: nil
\end{lstlisting}
This command selects which rules to apply and where according to a fixed criterion described in the Maude manual~\cite{maude}. This can be seen if the number of consecutive rewrites is limited to one with the \texttt{[1]} modifier.
\begin{lstlisting}[language={}]
Maude> rewrite [1] a b c .
rewrites: 1
result List: b c
\end{lstlisting}
The result is \texttt{b c} because the \texttt{pop} rule has been applied on the subterm \texttt{a}, extended by the identity axiom to \texttt{nil a}, but it could have also been applied on \texttt{(a b) c} yielding \texttt{a b} as a result. With the strategy language discussed in~\cref{sec:slang}, strategy modules (\kywd{smod}) can be used to specify different ways of applying these rules. 

	Maude specifications are accurately executable under certain requirements~\cite{maude}, like the confluence and termination of its equations and the coherence of its rules with them, since rule rewrites take place when the term is exhaustively reduced to a normal form with the oriented equations. Confluence and coherence can be easily defined for abstract reduction systems $(S, \to)$ where $S$ is a set of states and $(\to) \subseteq S \times S$ is a binary relation on them. This relation is \emph{confluent} if for any states $s, s_1, s_2 \in S$ satisfying $s \to^* s_1$ and $s \to^* s_2$, there is an $s' \in S$ such that $s_1 \to^* s'$ and $s_2 \to^* s'$, where $\to^*$ denotes the transitive and reflexive closure of $\to$. A state $s$ is \emph{irreducible} if there is no $s' \in S$ such that $s \to s'$, and we write $s \to^! s'$ if $s \to^* s'$ and $s'$ is irreducible. Whenever the relation is confluent and $s \to^! s'$, we say that $s'$ is a \emph{normal form} of $s$ and it is unique. Moreover, the relation is \emph{terminating} if there is no infinite sequence of states $(s_k)_{k=0}^\infty$ such that $s_k \to s_{k+1}$. For confluent and terminating relations, every state has a unique normal form. Given two relations $\to_1$ and $\to_2$ on $S$, $\to_2$ is \emph{coherent} with $\to_1$ if for any $s, s', u \in S$ such that $s \to_1^! u$ and $s \to_2 s'$, then there are $u', w \in S$ satisfying $u \to_2 u'$, $s' \to_1^! w$ and $u' \to_1^! w$. In other words, $\to_2$ is coherent with $\to_1$ if we compute steps of $\to_2$ modulo $\to_1$ by first reducing the state to its normal form by $\to_1$ and then applying $\to_2$.

\subsection{Reflection and metalevel computations} \label{sec:reflection}

	Rewriting logic is a reflective logic, whose objects and operations can be consistently represented in itself. Maude offers a predefined \emph{universal theory}~\cite[\S 17]{maude} to metatheoretically represent terms, equations, rules, modules, and so on. Operations like matching, reduction, and rule application can be programmed generically using regular operators and equations. However, Maude provides special operators backed by the object-level implementation in C++ to allow efficient reflective computations. Metarepresentations can in turn be metarepresented and terms be moved between different levels, thus yielding arbitrarily high reflective towers if needed.

	This universal theory is specified in \texttt{META-LEVEL} and its imported modules, and it relies on the \texttt{Qid} sort of \emph{quoted identifiers}, arbitrary words prefixed by an apostrophe. A variable \texttt{X} of sort \texttt{Nat} is metarepresented as the quoted identifier \texttt{'X:Nat}, and the constant \texttt{'Nat} of sort \texttt{Qid} is \texttt{''Nat.Qid}. Terms with arguments are represented using the operator \verb|_[_] : Qid NeTermList -> Term|, like \verb|'_+_['X:Nat, 's_['0.Zero]]| for \texttt{X + s~0}. Operator declarations, equations, rules, and similar statements are represented as terms with a syntax similar to the object-level reference. For example, the operator \texttt{+} may have a declaration \texttt{op '\_+\_ : 'Nat 'Nat -> 'Nat [comm assoc] .} and be involved in an equation \texttt{eq '\_+\_['X:Nat, '0.Zero] = 'X:Nat [none] .} where the trailing brackets enclose the set of operator or statement attributes, or \texttt{none} if there is none. Metamodules are terms with argument slots like \verb|fmod_is_sorts_.____endfm| for each kind of module component. Auxiliary functions \texttt{getOps}, \texttt{getEqs}, \texttt{getRls}, etc., are defined to obtain these components.

	Operations are accessible through some \emph{descent functions} like \texttt{metaMatch} for matching, \texttt{metaApply} for rule application, \texttt{metaReduce} for equational reduction, \texttt{metaRewrite}, etc. For instance, \texttt{metaReduce} receives the metarepresentations of a module and a term, and produces a pair containing the metarepresentations of the normal form of the given term and its calculated sort. Other predefined functions allow obtaining the metarepresentation of a term (\texttt{upTerm}) or the object-level term from its metarepresentation (\texttt{downTerm}). The metarepresentation of loaded modules can be obtained with \texttt{upModule} given the module name and a Boolean flag indicating whether a flat version (with all imports resolved) of the module is required, like in \texttt{upModule('NAT, false)}. The complete specification of the metalevel is in the Maude prelude and explained in~\cite[\S 17]{maude}.

\subsection{The Maude strategy language} \label{sec:slang}

	Strategies have been specified since the beginnings of Maude using the reflective features just explained. Nevertheless, to control rewriting in a more accessible and understandable way, an object-level strategy language was designed, based on that experience with reflective strategies and other strategy languages like ELAN~\cite{elan} and Stratego~\cite{stratego}. After being prototyped in Full Maude and tested, the language is finally implemented at the C++ level in Maude~3 with new features like compositional and parameterized strategy modules~\cite{pssm}. A strategy expression $\alpha$ in the language restricts the possible next steps during the rewriting process, and it can be seen as a transformation from an initial term $t$ to the set of terms that this controlled ---but not necessarily deterministic--- rewriting yields as a result. This is what the command \texttt{srewrite $t$ using $\alpha$} and its depth-first version \texttt{dsrewrite} show, by exploring all allowed execution paths.

	The application of a rule \texttt{$rl$[$x_1$ \textleftarrow{} $t_1$, $\ldots$, $x_n$ \textleftarrow{} $t_n$]\{$\alpha_1$, $\ldots$, $\alpha_m$\}} is the basic element of the strategy language, referred by its label $rl$ and taking an optional initial substitution. If the rule to be applied includes rewriting conditions, a comma-separated list of strategies must be given between curly brackets to control all of them. Rules are applied anywhere within the term by default, but their application can be restricted to the top with the \skywd{top}\texttt{($\alpha$)} combinator.
Moreover, the strategy \skywd{all} applies any labeled or unlabeled rule of the module with the \texttt{rewrite} command semantics.
Tests \skywd{match $P$ s.t.\ $C$} discard executions when the subject term does not match the pattern term $P$ or satisfy the equational condition $C$. The \skywd{match} keyword can be changed to \skywd{amatch} to match anywhere within the term, and the condition $\skywd{s.t.}\;C$ can be omitted if not needed. These elements can be combined with the concatenation $\alpha \texttt; \beta$ that executes $\beta$ on the results of $\alpha$, the disjunction $\alpha \texttt| \beta$ whose result comprises the executions allowed by any of its arguments, the iteration $\alpha \texttt*$ and normalization $\alpha \texttt!$ operators that iterate $\alpha$ any number of times or until no more iterations are possible, and the conditional $\alpha \texttt? \beta \texttt: \gamma$ that evaluates $\alpha$ and then $\beta$ on its results, but if $\alpha$ does not produce any, it executes $\gamma$ on the initial term. Two constants \skywd{idle} and \skywd{fail} represent the strategy that produces the initial term as result and the strategy that does not produce any result, respectively. The combinator \skywd{matchrew $P$ s.t. $C$ by $x_1$ using $\alpha_1$, $\ldots$, $x_n$ using $\alpha_n$} allows rewriting selected subterms of the term where it is applied: those terms matched by the variables $x_1, \ldots, x_n$ in the pattern are rewritten in parallel using $\alpha_1, \ldots, \alpha_n$, respectively, and their results replace the matched subterms to produce the \skywd{matchrew} results. This initial keyword can also be changed to \skywd{amatchrew} like for tests. Moreover, for efficiency reasons, the combinator \skywd{one}\texttt{($\alpha$)} evaluates $\alpha$ only until the first solution is found, if any, which is returned as its result. In addition to these core combinators, the language includes some others that can be defined in terms of the first, like \texttt{\skywd{try}($\alpha$) $\equiv$ $\alpha$ ? \skywd{idle} : \skywd{idle}}, \texttt{\skywd{not}($\alpha$) $\equiv$ $\alpha$ ? \skywd{fail} : \skywd{idle}}, and \texttt{\skywd{test}($\alpha$) $\equiv$ \skywd{not}(\skywd{not}($\alpha$)}.  For example, using the \texttt{LLIST-M} module at the beginning of this section, we can execute the following strategy:
\begin{lstlisting}[language={}]
Maude> srewrite a b c using top(pop) ; top(put[L <- d]) .

Solution 1
rewrites: 2
result List: a b d

No more solutions.
\end{lstlisting}
Remember that \texttt{L} is the name of the unbounded variable of the \texttt{put} rule.

	Strategy modules allow declaring and defining strategies with a name and any number of arguments. Delimited by the \kywd{smod} and \kywd{endsm} keywords, they may import modules of any kind and include any statement available in functional and system modules, although a clear separation of the model from its control encourages that only strategy declaration and definitions statements are used. Named strategies can be declared as \lstinline[mathescape, basicstyle=\ttfamily]|strat $\mathit{name}$ : $s_1 \;\cdots\; s_n$ @ $s$ .| with the signature of its arguments and the sort $s$ of the terms to which it is intended to be applied. They are defined with \lstinline[mathescape, basicstyle=\ttfamily]|sd $\mathit{name}$($p_1$, $\ldots$, $p_n$) := $\alpha$ .| or \lstinline[mathescape, basicstyle=\ttfamily]|csd $\mathit{name}$($p_1$, $\ldots$, $p_n$) := $\alpha$ if $C$.| if they are conditional, which are only executed when their equational condition $C$ is satisfied. Strategies can be called even recursively in strategy expressions as \texttt{$\mathit{name}$($t_1$, $\cdots$, $t_n$)}. Extending the example module \texttt{LLIST-M} again, a strategy \texttt{seq} is defined to append a list of letters to the list on which it is applied:
\begin{lstlisting}
smod LLIST is
  protecting LLIST-M .
  var LS : List . vars L L' : Letter .

  strat seq : List @ List .
  sd seq(nil)  := idle .
  sd seq(L' LS) := top(put[L <- L']) ; seq(LS) .
endsm
\end{lstlisting}
Note that \texttt{L} in \texttt{put[L <- L']} refers to the variable \texttt{L} in the rule that is being instantiated, while \texttt{L'} refers to the strategy argument that decides its value.

	The strategy language and strategy modules are also represented at the metalevel, faithfully reproducing the object-level syntax in most cases. Its combinators are specified as terms of the \texttt{Strategy} sort in the \texttt{META-STRATEGY} module. For instance, a simple rule application is denoted as \verb|'label[none]{empty}| and a strategy call as \texttt{'name[[$\mathit{TL}$]]} with $\mathit{TL}$ a possibly \texttt{empty} list of metarepresented terms. Strategy modules
\begin{lstlisting}
op smod_is_sorts_._______endsm : Header ImportList SortSet
  SubsortDeclSet OpDeclSet MembAxSet EquationSet RuleSet
  StratDeclSet StratDefSet -> StratModule [ctor ...] .
\end{lstlisting}
as well as strategy declarations and definitions
\begin{lstlisting}
op sd_:=_[_]. : CallStrategy Strategy AttrSet 
                 -> StratDefinition [ctor ...] .
op csd_:=_if_[_]. : CallStrategy Strategy EqCondition
             AttrSet -> StratDefinition [ctor ...] .
\end{lstlisting}
are specified too, and the commands \texttt{srewrite} and \texttt{dsrewrite} are accessible through the \texttt{metaSrewrite} descent function.
\begin{lstlisting}
op metaSrewrite : Module Term Strategy SrewriteOption Nat
                   ~> ResultPair? [special (...)] .
sort SrewriteOption .
ops breadthFirst depthFirst : -> SrewriteOption [ctor] .
\end{lstlisting}
The last argument of the \texttt{metaSrewrite} operator is an index used to enumerate the potentially multiple solutions, until a \texttt{failure} term is obtained. These solutions are provided as pairs \texttt{\{\_,\_\}} of sort \texttt{ResultPair} containing the metarepresentation of the term and its calculated sort. Following a common notational pattern in Maude, the sort \texttt{ResultPair?}\ designates a supersort of \texttt{ResultPair} with the additional constant \texttt{failure} to indicate the absence of a result, as explained before.

	Another useful descent function for building metalanguage interfaces is \texttt{metaParse} that parses terms on a given module and sort.
\begin{lstlisting}
op metaParse : Module VariableSet QidList Type? 
               ~> ResultPair? [special (...)] .
\end{lstlisting}
On success, it returns a pair with the metarepresentation of the parsed term and its least sort from a list of tokens of sort \texttt{QidList}, which can be obtained from a string using the \texttt{tokenize} function.

	In previous prototypes of the Maude strategy language there was nothing like a metalevel of the strategy language, since it was specified within Maude and strategy expressions were directly Maude terms. The language was more easily extensible at the expense of efficiency, since the execution of strategies was implemented in Maude itself.

	More details on the language can be found in~\cite[\S 10]{maude}.

\subsection{Interactive interfaces}

	Writing interactive interfaces in Maude is relatively easy, and it is usually done to offer a convenient interface to the logic and semantic frameworks specified in the language. The archetype is Full Maude~\cite[\S 15]{maude}, an extended interpreter written in Maude itself where many features later implemented in C++ have been first tested. The functionality of the Core Maude interpreter is replicated there along with additional features like tuple types and object-oriented modules. Users can also extend Full Maude to include their own features and commands.
Moreover, since Maude~3~\cite{maude30}, the interactive capabilities of Maude have increased due to new \emph{external objects} that allow reading and writing files as well as the standard input and output streams. External objects are an object-oriented mechanism that allows Maude programs to communicate with the outside world, already used in previous versions for Internet sockets. The standard \texttt{CONFIGURATION} module defines an extensible signature for defining objects and messages, which are held in a common soup or multiset where objects read and introduce messages by means of rewriting rules. The command \texttt{erewrite} conducts rewriting of these configurations following an object-fair strategy and handling the messages issued to and by the implicit external objects. In this case, the \texttt{STD-STREAM} module in the \texttt{file.maude} file of the Maude distribution declares the \texttt{stdin} and \texttt{stdout} objects, and the \texttt{getLine}/\texttt{gotLine} and \texttt{write}/\texttt{wrote} messages to read and write to the terminal.

\subsection{Model checking} \label{sec:modelchecking}

	Model checking is an automated verification technique that explores all possible executions of a system to verify whether it meets a given specification. This umbrella term comprises different algorithms and multiple variations, but its models are essentially based on annotated transition systems $\mathcal K = (S, \to, I, AP, \ell)$ known as Kripke structures, where $\to$ is a (sometimes labeled) binary relation and $AP$ is a finite set of atomic propositions associated to each state by a labeling function $\ell : S \to \mathcal P(AP)$. Properties are usually expressed in terms of these atomic propositions (and possibly on the labels of its transitions) using some temporal logics that include temporal operators describing how they occur in time. Examples of well-known logics are LTL~\cite{pneuliLTL}, CTL~\cite{ctl}, their superset CTL*~\cite{ctlstar}, and $\mu$-calculus~\cite{mucalcmc}.

	Rewriting systems can be naturally seen as Kripke structures whose states are terms and whose transitions are one-step rule rewrites. Maude specifications can be model checked against LTL properties since its 2.0 version thanks to a builtin model checker~\cite[\S 12]{maudemc,maude}. We have extended it for systems controlled by strategies~\cite{fscd}, and for the other logics mentioned in the previous paragraph~\cite{btimemc}. Since strategies describe a subset or subtree of allowed executions of the model, properties are satisfied by a strategy-controlled model iff they are satisfied on this subset or subtree. The only question remaining is which are the executions described by a Maude strategy language expression. This is answered by a small-step operational semantics, respected by the model checker implementations. For checking properties other that LTL, some external model checkers can be used, including LTSmin~\cite{LTSmin}, through an extensible model-checking interface \textsf{umaudemc}~\cite{umaudemc} that unifies the interaction and the syntax of the logics. This interface is built over a library that allows accessing Maude objects and operations from Python and other programming languages. We use this library to adapt \textsf{umaudemc} for the various examples in this paper.

\section{An introductory example \refcd{http://maude.ucm.es/strategies/examples/munorm.maude}} \label{sec:1stexample}

	This section is intended as a preface through a simple example to the metaprogramming resources offered by Maude to manipulate and generate strategies from some data, like the metarepresentation of a Maude module. We will follow the same method that is applied to the more complex examples of the following sections and that can be applied in general for other reflective transformations. In this example, a strategy will be generated to normalize terms while respecting certain constraints that are included in the specification of a rewriting or functional program.

\begin{figure}\centering
\begin{tikzpicture}[level distance=2em, sibling distance=3em]
\node (T0) {\ttfamily\color{blue} \underline{take}}
	child {node {\ttfamily 2}}
	child {node {\ttfamily natsFrom}
		child {node {\ttfamily 1}}};

\node[right=1.9cm of T0] (T1) {\ttfamily :}
	child {node {\ttfamily \color{blue} \underline{1}}}
	child {node {\ttfamily \color{blue} \underline{take}}
		child {node {\ttfamily 1}}
		child {node {\ttfamily natsFrom}
			child {node {\ttfamily +}
				child {node {\ttfamily 1}}
				child {node {\ttfamily 1}}
			}
		}
	};

\node[right=2.7cm of T1] (T2) {\ttfamily :}
	child {node {\ttfamily 1}}
	child {node {\ttfamily :}
		child {node {\ttfamily \color{blue} \underline{2}}}
		child {node {\ttfamily \color{blue} \underline{take}}
			child {node {\ttfamily 0}}
			child {node {\ttfamily natsFrom}
				child {node {\ttfamily +}
					child {node {\ttfamily 2}}
					child {node {\ttfamily 1}}
				}
			}
		}
	};

\node[right=3cm of T2] (T3) {\ttfamily :}
	child {node {\ttfamily 1}}
	child {node {\ttfamily :}
		child {node {\ttfamily 2}}
		child {node {\ttfamily nil}}
	};

\draw[->] (-.2, -2.8) -- node[above] {$\mu$-normalization} (1.3, -2.8);

\end{tikzpicture}
\caption{Layered normalization of \texttt{take(2, natsFrom(1))}} \label{fig:layered}
\end{figure}
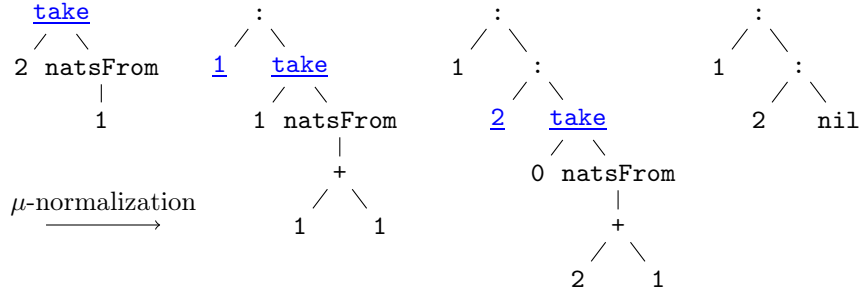

	\emph{Context-sensitive rewriting}~\cite{contextRew} is a restricted form of term rewriting defined by simple constraints attached to the symbols of the signature that exclude some of their arguments from being rewritten. Maude has builtin support for this kind of restrictions by means of the \texttt{strat} and \texttt{frozen} attributes.
\begin{lstlisting}[mathescape, moredelim={[is][]{\#}{\#}}]
op $f$ : $s_1$ $\cdots$ $s_n$ -> $s$ [#strat#($i_1$ $\cdots$ $i_k$ 0) frozen($j_1$ $\cdots$ $j_l$)] .
\end{lstlisting}
Regarding equational reduction, the evaluation strategy attribute \texttt{strat} specifies a zero-terminated list of argument indices $i_m \in \{ 1, \ldots, n \}$ that fix the order in which arguments are reduced before applying equations to the top, while absent arguments are not reduced at all. By default, the evaluation strategy is \texttt{1 2 $\cdots$ $n$ 0}. Regarding rules, the \texttt{frozen} attribute inhibits rewriting with rules inside a given subset of arguments $j_m \in \{1, \ldots, n\}$.
These restrictions may prevent non-terminating evaluations, but their direct application is not enough to obtain irreducible terms, for which strategies are needed, as we will see with a lazy programming example. Generating these strategies from the context-sensitive restrictions is the purpose of our introductory metalevel transformation. Let us present first the following functional module~\cite{csrWrla04} that attempts to specify a lazy list of integers:\footnote{Natural numbers are represented in Maude using Peano notation with a constant \texttt{0} and a successor operator \texttt{s\_}, although numeric literals can also be written as syntactic sugar. Integers include an additional constructor \texttt{-\_}.}
\begin{lstlisting}
fmod LAZY-LIST is
  protecting INT .
  sort LazyList .

  op nil : -> LazyList [ctor] .
  op _:_ : Int LazyList -> LazyList [ctor] .

  var E : Int . var N : Nat . var L : LazyList .

  op take : Nat LazyList -> LazyList .
  eq take(0, L) = nil .
  eq take(s(N), E : L) = E : take(N, L) .

  op natsFrom : Nat -> LazyList .
  eq natsFrom(N) = N : natsFrom(N + 1) .
endfm
\end{lstlisting}
Even though \texttt{natsFrom($n$)} represents an infinite list, containing all natural numbers from $n$, we would expect that the lazy evaluation of a term like \texttt{take(3, natsFrom(0))} leads to \texttt{0:1:2:nil}. However, Maude's \texttt{reduce} command eagerly applies equations in an innermost leftmost manner, so the evaluation of this term will not terminate because of the continuous reduction of the tails in the \texttt{natsFrom} definition.
Fortunately, the Maude \texttt{strat} attribute can be used on the \verb|_:_| operator to avoid reducing the tail of the list, by changing its \texttt{[ctor]} attribute to \texttt{[ctor strat(1 0)]}. However, this context-sensitive restriction limits rewriting too much, and no valid result is still produced:
\begin{lstlisting}
Maude> reduce take(3, natsFrom(0)) .
rewrites: 2
result LazyList: 0 : take(2, natsFrom(0 + 1))
\end{lstlisting}

In the vocabulary of context-sensitive rewriting, \texttt{strat} and \texttt{frozen} annotations correspond to \emph{replacement maps} $\mu : \Sigma \to \mathcal P(\N)$ where $\mu(f) \subseteq \{1, \ldots, \mathrm{arity}(f) \}$ for all $f \in \Sigma$. Reduction is only allowed in the \emph{$\mu$-replacing positions} of any term, defined recursively as
\[ \mathrm{Pos}^\mu(f(t_1, \ldots, t_n)) = \{\varepsilon\} \cup \bigcup_{i \in \mu(f)} \{i\} \, \mathrm{Pos}^\mu(t_i), \]
where $\varepsilon$ denotes the top position and the word $w i$ the $i$-th argument of the subterm at position $w$. Exhaustively reducing in these positions yields \emph{$\mu$-normal forms}, exactly what the previous command did for \texttt{take(3, natsFrom(0))} and $\mu(\texttt{\_:\_}) = \{1\}$. As that execution shows, $\mu$-normal forms are not necessarily normal forms of the unrestricted rewrite system, but $\mu$-normalization can be useful to build complete and lazy normalization procedures~\cite{csrWrla04}. Normalization can be achieved via $\mu$-normalization using a \emph{layered evaluation} that safely resumes reduction on the subterms of non-replacing positions~\cite[\S 9.3]{contextRew}, as illustrated in~\cref{fig:layered} for the term \texttt{take(2, natsFrom(1))}. At each step, the highlighted subterms at a given level of the term tree are applied $\mu$-normalization, and this continues to its arguments down to the leaves. This procedure is implemented by means of a strategy proposed by Salvador Lucas~\cite{appContextSens}, which would need to traverse the term. Since the Maude strategy language does not offer any resource to do it generically, a signature-aware strategy must be produced.

	The following function \texttt{csrTransform} implements a metalevel module transformation that extends the metarepresentation of the input module \texttt{M} with strategy declarations and definitions that normalize terms as described in the previous paragraph. Its global shape is given by the following equation.\footnote{Strategy declarations must include the intended sort to which they are applied after the \texttt{@} sign, although this is merely informative and reasonable candidates do not always exist. Since the strategies defined here are somehow polymorphic, we declare \texttt{AnyTerm} just to take its place.}
\begin{lstlisting}[escapechar=^, keywordstyle={[2]{}}]
op csrTransform : Module -> StratModule .
eq csrTransform(M) = smod append(getName(M), 'CSR)  is
  getImports(M)  ^\hfill^ *** module importation
  sorts 'AnyTerm ; getSorts(M) .  ^\hfill^ *** sort decls
  getSubsorts(M) ^\hfill^ *** subsort decls
  strat2frozen(getOps(M)) ^\hfill^ *** operator decls
  getMbs(M) ^\hfill^ *** sort membership axioms
  none ^\hfill^ *** equations
  getRls(M) ^\hfill^ *** rules
  eqs2rls(getEqs(M))
  getStrats(M) ^\hfill^ *** strategy decls
  (strat 'norm-via-munorm : nil @ 'AnyTerm [none] .)
  (strat 'munorm : nil @ 'AnyTerm [none] .)
  (strat 'decomp : nil @ 'AnyTerm [none] .)
  getSds(M) ^\hfill^ *** strategy definitions
  (sd 'norm-via-munorm[[empty]] :=
       'munorm[[empty]] ; 'decomp[[empty]] [none] .)
  (sd 'munorm[[empty]] := one(all) ! [none] .)
  (sd 'decomp[[empty]] := makeDecomp(getOps(M)) [none] .)
endsm .
\end{lstlisting}
Except for the new strategies, the transformed module is essentially a copy of the original one. However, since the Maude strategy language can only control rule application, we translate all equations into rules,\footnote{Equations in Maude can be annotated with the \texttt{owise} attribute that cause them to be executed at a certain position only if equations without this attribute have failed. Respecting the \texttt{owise} semantics would require specifying strategies to apply their translation as rules likewise, but for simplicity we assume that there are no \texttt{owise} annotations.} 
and all \texttt{strat} attributes to \texttt{frozen} annotations.
\begin{lstlisting}[moredelim={[is][]{\#}{\#}}]
eq eqs2rls(none) = none .
eq eqs2rls(#eq# L =  R [Attrs] . Eqs) =
           #rl# L => R [Attrs] . eqs2rls(Eqs) .
eq eqs2rls(#ceq# L =  R #if# C [Attrs] . Eqs) = 
           #crl# L => R #if# C [Attrs] . eqs2rls(Eqs) .
\end{lstlisting}
The transformed module is always a strategy module, regardless of which type of module \texttt{M} is, where three strategies are declared. The entry point for the layered normalization procedure is \texttt{norm-via-munorm}, which executes two auxiliary strategies \texttt{munorm} for $\mu$-normalization, and then \texttt{decomp} for resuming normalization inside frozen arguments. \texttt{munorm} is implemented by exhaustively (\texttt{!}) applying the rules in the module respecting the \texttt{frozen} restrictions (\skywd{all}). Assuming the input system is $\mu$-confluent, i.e.\ confluent under the context-sensitive restrictions, the order in which rules are applied does not affect the result, so \skywd{all} is executed for efficiency under the \skywd{one} operator that discards alternative rewrite orders. For its part, the \texttt{decomp} strategy continues normalization on the symbol arguments. Strategies can be applied inside subterms in the Maude strategy language using the \skywd{matchrew} combinator, so one is generated for each $f \in \Sigma$ to apply \texttt{norm-via-munorm} recursively to every subterm:
\[ \texttt{\skywd{matchrew} $f$($x_1$, $\ldots$, $x_n$) \skywd{by} $\ldots$, $x_i$ \skywd{using} norm-via-munorm, $\ldots$} \]
The decomposition strategy \texttt{decomp} is defined as the disjunction of all these combinators. Only the one for the top symbol of the term where it is applied on each occasion will match. Since this definition depends on the signature of the module, \texttt{decomp} is reflectively generated by the \texttt{makeDecomp} function that walks through the operators declared in the module.
\begin{lstlisting}[keywordstyle={[2]{}}, escapechar=^]
var Q  : Qid .    var Ops   : OpDeclSet .    var N : Nat .
var Ty : Type .   var NeTyL : NeTypeList .

op makeDecomp : OpDeclSet -> Strategy .
eq makeDecomp(none) = fail .
eq makeDecomp(^op^ Q : nil -> Ty [Attrs] . Ops) =
     (match qid(string(Q) + "." + string(Ty)) s.t. nil) 
     | makeDecomp(Ops) .
eq makeDecomp(^op^ Q : NeTyL -> Ty [Attrs] . Ops) =
  (matchrew Q[makeVarList(NeTyL, 1)] s.t. nil
    by makeUsingPart(NeTyL, 1)) | makeDecomp(Ops) .
\end{lstlisting}
Constants (operators with an empty list \texttt{nil} of arguments) do not have arguments in which normalization should be resumed, but they must also be matched by the \texttt{decomp} strategy so that it does not fail when any of them is encountered.\footnote{Instead of including constants in the disjunction of the \texttt{decomp} strategy, we could have surrounded the call to \texttt{decomp} with the \texttt{try} combinator so that it does not fail when no pattern matches.} In this case, instead of a \skywd{matchrew}, a test \skywd{match} is used with the metarepresentation of that constant as described in~\cref{sec:reflection}. Auxiliary functions like \texttt{makeVar} and \texttt{makeVarList} are used to generate sequentially-numbered variable metarepresentations of the given sorts for the \skywd{matchrew} pattern. These variables are mapped to the \texttt{norm-via-munorm} strategy by the \texttt{makeUsingPart} function.
\begin{lstlisting}[keywordstyle={[2]{}}]
op makeUsingPart : NeTypeList Nat -> UsingPairSet .
eq makeUsingPart(Ty, N) =
     makeVar(N, Ty) using 'norm-via-munorm[[empty]] .
eq makeUsingPart(Ty NeTyL, N) = makeUsingPart(Ty, N),
                                makeUsingPart(NeTyL, s(N)) .

op makeVar : Nat Type -> Variable .
eq makeVar(N, Ty) =
     qid("X" + string(N, 10) + ":" + string(Ty)) .
\end{lstlisting}

	Finally, the term \texttt{csrTransform(upModule('LAZY-LIST, true))} can be reduced to obtain the transformed \texttt{'LAZY-LIST} module. Remember that \texttt{upModule(\linebreak'$\mathit{name}$, true)} evaluates to the flat metarepresentation of the module $\mathit{name}$ where all importations have been resolved.
Then, the \texttt{norm-via-munorm} strategy can be applied to a term using the \texttt{metaSrewrite} function, whose inputs and results are written at the metalevel:
\begin{lstlisting}[language={}]
red metaSrewrite(csrTransform(upModule('LAZY-LIST, true)),
                 'take['s_^3['0.Zero], 'natsFrom['0.Zero]],
                 'norm-via-munorm[[empty]],
                 breadthFirst, 0) .
rewrites: 3123
result ResultPair: {'_:_['0.Zero,'_:_['s_['0.Zero],
            '_:_['s_^2['0.Zero],'nil.LazyList]]],'LazyList}
\end{lstlisting}
Alternatively, Full Maude can be used with terms and strategies at the object level:\footnote{Full Maude commands are typed between parentheses, once the \texttt{full-maude.maude} file is loaded. Its latest version can be downloaded from \href{http://maude.cs.illinois.edu}{\ttfamily maude.cs.illinois.edu}.}
\begin{lstlisting}[language={}]
(select CSR-TRANSFORM .)
(load csrTransform(upModule('LAZY-LIST, true)) .)
(select LAZY-LIST-CSR .)
(srewrite take(3, natsFrom(0)) using norm-via-munorm .)

Solution 1
result LazyList: 0 : 1 : 2 : nil

No more solutions
\end{lstlisting}
The evaluation now terminates with a meaningful result. When the rewrite system is confluent and terminating under the restrictions, as in this case, any search strategy, \texttt{breadthFirst} or \texttt{depthFirst}, \texttt{srewrite} or \texttt{dsrewrite}, would produce the same result since there is a single solution and a finite state space. The \texttt{norm-via-munorm} strategy can be used to normalize any term in the module without computing the transformation again. However, since it explicitly refers to the signature of that module, its definition is not applicable to any other module.

	The following section shows an extension of the Maude strategy language adding new combinators based on the subject module in which strategies are to be applied. Using that extended language the \texttt{norm-via-munorm} strategy will be defined directly and more succinctly.

\section{Theory-dependent extensions of the strategy language \refcd{http://maude.ucm.es/strategies/examples/congruenceOpsExt.maude}} \label{sec:langext}

	When the Maude strategy language was designed, the objective was not to offer a vast repertory of operators to concisely express a wide range of tasks, like in the case of Stratego~\cite{stratego}, but to be compact and expressive enough. Thanks to reflection, the language can be extended to better suit a specific purpose or to incorporate a new feature. In this section, we apply this principle and describe a general schema to construct strategy language extensions by some module transformations without losing any of the advantages of the strategy language, like the interaction at the object level and with the strategy-aware model checker. In particular, the language is extended with the so-called \emph{congruence operators} from ELAN~\cite{elan} and Stratego, and the \emph{generic traversals} from Stratego. Both operator families depend on the signature of the subject module where strategies are applied, so a module transformation is required to implement them.
	
\begin{figure}[t]\centering
\begin{tikzpicture}
\node[draw] (ML) {\ttfamily META-LEVEL};
	\node[draw, right=2cm of ML] (SG) {\ttfamily SLANG-GRAMMAR};
	\node[draw, below=of ML] (E) {$\mathit{ext}(M)$};
	\node[draw, below=of SG] (ISG) {$\mathit{extGram}(M)$};

	\coordinate (XM) at ($(ML.east)!0.5!(SG.west)$);
	\coordinate (YM) at ($(ML.south)!0.5!(E.north)$);
	\node[draw] (M) at (XM |- YM) {$M$};

	\draw[->] (ML) -- (E);
	\draw[->] (SG) -- (ISG);
	\draw[->, dashed] (M) -- ($(ML.south)!0.5!(E.north)$);
	\draw[->, dashed] (M) -- ($(SG.south)!0.5!(ISG.north)$);

\node[below=1ex of ISG] (T) {$t$};
	\node[below=1ex of E] (A) {$\overline\alpha$};
	\node[left=1ex of ML] (B) {$\overline\beta$};

	\node[right=of T] {\ttfamily\small metaParse};
	\node[left=of B] {\ttfamily\small metaSrewrite};

	\draw[->] (T) edge node[above] {stratParse} (A);
	\draw (A) -- (B |- A) edge[->] node[left] {transform} (B);
\end{tikzpicture}
\caption{Typical structure of a strategy language extension} \label{fig:extension}
\end{figure}
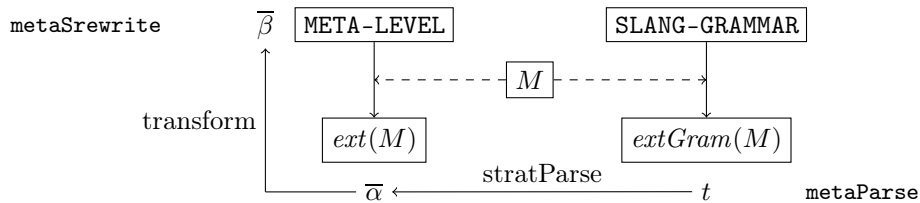

	The procedure we will follow is directly applicable to other extensions, and it is supported by a collection of helper modules that save work to the extension developer and avoid writing boilerplate code for each extension. It consists of the following steps illustrated in~\cref{fig:extension}:
\begin{enumerate}
	\item Extending the universal theory of the \texttt{META-LEVEL} module with the metarepresentation of the new strategy combinators, probably depending on the module $M$ where strategies are to be applied. This is similar to what we did in~\cref{sec:1stexample}.
	\item Since the builtin \texttt{metaSrewrite} function does not support the new combinators and in order to execute them, extended expressions are translated to the standard language by extending the skeleton of a \texttt{transform} function. Moreover, the translation allows using the strategy-aware model checker on extended strategies and all other strategy-related machinery of the interpreter for free. \item In order to write extended strategies at the object level, an extensible grammar \texttt{SLANG-GRAMMAR} of the strategy language can be added productions for the new operators. Strategies are parsed as terms with the builtin \texttt{metaParse} function, and transformed to their metarepresentations by an extensible \texttt{stratParse} function.
	\item A parameterized interactive Maude interface is provided to operate with extended strategies completely at the object level. It admits extended strategies in its \texttt{srewrite} command and in the strategy definitions of strategy modules. A \texttt{umaudemc}-based program for model checking LTL, CTL*, and $\mu$-calculus with these strategies and modules is available too (see~\cref{sec:modelchecking}).
\end{enumerate}
As a result, extended strategy expressions can be used almost anywhere an original expression could have been used, although not directly in the commands of the Maude interpreter. Since strategies are translated to the standard strategy language, some extensions are not easily implemented using this approach, as we discuss at the end of the section.

	As said, we will exemplify this procedure with two families of operators that are not available in the Maude language. Congruence operators $f(\alpha_1, \ldots, \alpha_n)$ are strategy combinators that reproduce the data constructors of the target module $f(t_1, \ldots, t_n)$ with their arguments replaced by strategies, which are applied to the corresponding arguments of the subject term's top symbol if they coincide. In other words, they are semantically equivalent to a \skywd{matchrew} construct of the form
\[ \texttt{$f$($\alpha_1$, $\ldots$, $\alpha_n$) $\equiv$ \skywd{matchrew} $f$($x_1$, $\ldots$, $x_n$) \skywd{by} $x_1$ \skywd{using} $\alpha_1$, $\ldots$, $x_n$ \skywd{using} $\alpha_n$}. \]
On the other hand, generic traversals are operators that allow applying a strategy along the structure of any term without explicitly mentioning it: \skywd{gt-all}\texttt{($\alpha$)} applies $\alpha$ to all arguments of the top symbol, \skywd{gt-one}\texttt{($\alpha$)} applies $\alpha$ to the first argument from left to right in which it succeeds, and \skywd{gt-some}\texttt{($\alpha$)} applies $\alpha$ to as many children as possible and at least to one, so it is equivalent to \texttt{\skywd{test}(\skywd{gt-one}($\alpha$)) ; \skywd{gt-all}(\skywd{try}($\alpha$))}.\footnote{The original names of generic traversal operators in Stratego do not include the \texttt{gt-} prefix, which is used here to avoid confusion with the \skywd{one} and \skywd{all} operators of the strategy language.}

	\paragraph{Congruence operators} First, we extend the metalevel with the metarepresentations of congruence operators: a homonym symbol of sort \texttt{Strategy} is introduced for each data constructor of $M$ taking as many \texttt{Strategy} arguments as the arity of the original one:
\begin{lstlisting}[moredelim={[is][]{\#}{\#}}]
op generateCongOps : OpDeclSet -> OpDeclSet .
eq generateCongOps(none) = none .
eq generateCongOps(#op# Q : TyL -> Ty [ctor Attrs] . Ops) =
  (#op# Q : repeatType('Strategy, size(TyL))
           -> 'Strategy [ctor removeId(Attrs)] .)
  generateCongOps(Ops) .
eq generateCongOps(Op Ops) = generateCongOps(Ops) [owise] .
\end{lstlisting}
The auxiliary function \texttt{repeatType} builds a list with the given number of repetitions of its first argument, and \texttt{removeId} removes identity axiom attributes of the original operators, which are meaningless in the congruence operators. Overloaded symbols may produce different conflicting declarations if their attributes do not coincide, so generated operators are given a second pass to remove potential conflicts.

	Extended strategies can now be expressed at the metalevel, but since the builtin \texttt{metaSrewrite} is unaware of these new operators and we do not want to implement the strategy language from scratch, we should translate them to the standard subset. This translation is defined in the extended \texttt{META-LEVEL} as a function \texttt{transform} between terms of sort \texttt{Strategy}. The complete recursive definition of this function would be large and repetitive, so an extendable and generic one is supplied to facilitate the task of defining extensions. This is provided by the \texttt{SLANG-EXTENSION-STATIC} module to be included in the transformed module, where equations are given for the standard constructors and the user only has to provide equations for the new elements.
\begin{lstlisting}[escapechar=^,keywordstyle={[2]{}}]
fmod SLANG-EXTENSION-STATIC is
  protecting META-LEVEL .

  op transform : Strategy Nat -> Strategy .

  var N : Nat .      var S : Strategy .
  eq transform(idle, N) = idle .
  eq transform(top(S), N) = top(transform(S, N)) .
  ^\ldots^
endfm
\end{lstlisting}
The \texttt{transform} operator takes a natural number as a second argument, used as an index to generate fresh variables in nested \skywd{matchrew}s, since their bindings are permanent. Some helper functions like \texttt{makeVar} and \texttt{makeConstant} that already appear in the previous example are included in the module too. Note that the equations defining \texttt{transform} are generated by the module transformation, so they must metarepresent \texttt{Strategy} terms and involve two levels of reflection.
\begin{lstlisting}[moredelim={[is][]{\#}{\#}}]
op generateCongOpsDefs : OpDeclSet -> EquationSet .

eq generateCongOpsDefs(none) = none .
eq generateCongOpsDefs(#op# Q : NeTyL -> Ty [ctor Attrs] . Ops) =
  (#eq# 'transform[Q[makeStratVars(size(NeTyL))], 'N:Nat] =
    'matchrew_s.t._by_[
      '_`[_`][upTerm(Q), 'makeOpVars[upTerm(NeTyL), 'N:Nat]],
        'nil.EqCondition,
        'makeUsingPairs[upTerm(NeTyL),
           wrapStratList(makeStratVars(size(NeTyL))),
           '_+_['N:Nat, upTerm(size(NeTyL))], 'N:Nat]
    ] [none] .)
  generateCongOpsDefs(Ops) .

eq generateCongOpsDefs(#op# Q : nil -> Ty [ctor Attrs] . Ops) =
  (#eq# 'transform[qid(string(Q) + ".Strategy"), 'N:Nat] =
    'match_s.t._[makeConstant(Q, Ty), 'nil.EqCondition] [none] .)
  generateCongOpsDefs(Ops) .
eq generateCongOpsDefs(Op Ops) = 
     generateCongOpsDefs(Ops) [owise] .
\end{lstlisting}
The second equation generates the \skywd{matchrew} constructs described at the beginning of the section, and again, simpler \skywd{match} tests are used for constants. Variable names are generated from the index passed as the second argument of \texttt{transform}, which is increased in recursive calls to ensure that the index is not used again in a subterm. Just like when declaring them, the same congruence operator may receive multiple \texttt{transform} equations for different overloaded data constructors, so they are combined afterwards in a strategy disjunction.
\begin{lstlisting}[moredelim={[is][]{\#}{\#}}]
eq combineCongOpsDefs(Eqs
    (#eq# 'transform[T, 'N:Nat] = T1 [none] .)
    (#eq# 'transform[T, 'N:Nat] = T2 [none] .)) =
  combineCongOpsDefs(Eqs
     #eq# 'transform[T, 'N:Nat] = '_|_[T1, T2] [none] .) .
eq combineCongOpsDefs(Eqs) = Eqs [owise] .
\end{lstlisting}

	\paragraph{Generic traversals} Since the strategy language does not provide the means to perform generic traversals of terms, and since we have chosen to translate extended strategies to standard ones, we should implement generic traversals using module-specific strategies. Namely, we can translate the strategy \skywd{gt-all}\texttt{($\alpha$)} to the disjunction of \texttt{$f$($\alpha$, $\ldots$, $\alpha$)} for all $f \in \Sigma$, and \skywd{gt-one}\texttt{($\alpha$)} using the disjunction for all $f$ of
\begin{lstlisting}[mathescape]
$f$($\alpha$, idle, $\ldots$, idle) or-else $\cdots$ or-else $f$(idle, idle, $\ldots$, $\alpha$).
\end{lstlisting}
These still extended strategies are translated to the standard language as explained before. For instance, the strategies in the disjunction to which \skywd{gt-all} is translated can be built with the following equations:
\begin{lstlisting}[moredelim={[is][]{\#}{\#}}]
op generateGTAll : OpDeclSet -> TermList .

eq generateGTAll(none) = empty .
eq generateGTAll(#op# Q : NeTyL -> Ty [ctor Attrs] . Ops) =
  Q[repeatTerm('S:Strategy, size(NeTyL))],
  generateGTAll(Ops) .
eq generateGTAll(#op# Q : nil -> Ty [ctor Attrs] . Ops) =
  makeConstant(Q, 'Strategy), generateGTAll(Ops) .
eq generateGTAll(Op Ops) = generateGTAll(Ops) [owise] .
\end{lstlisting}
The final shape of the extended metalevel with congruence operators and generic traversals is given by the equation below. Generic traversals are defined directly with equations instead of using \texttt{transform}, because they do not explicitly mention any variable indices.
\begin{lstlisting}[moredelim={[is][]{\#}{\#}}]
eq extendCongOps(M) = fmod append('META-LEVEL, getName(M)) is
  (extending 'SLANG-EXTENSION-STATIC .)
  sorts none .
  none	*** subsorts
  combineCongOps(generateCongOps(getOps(M)))
  (op '#gt-all#  : 'Strategy -> 'Strategy [none] .)
  (op 'gt-one  : 'Strategy -> 'Strategy [none] .)
  (op '#gt-some# : 'Strategy -> 'Strategy [none] .)
  none	*** membership axioms
  combineCongOpsDefs(generateCongOpsDefs(getOps(M)))
  (eq '#gt-all#['S:Strategy] =
         '_|_[generateGTAll(getOps(M))] [none] .)
  (eq 'gt-one['S:Strategy] =
         '_|_[generateGTOne(getOps(M))] [none] .)
  (eq '#gt-some#['S:Strategy] =
         '_|_[generateGTSome(getOps(M))] [none] .)
endfm .
\end{lstlisting}
The \texttt{META-LEVEL} module is imported transitively via the already-known module \texttt{SLANG-EXTENSION-STATIC}. The complete specification is available in the strategy language example collection~\cite{stratweb}.

	\paragraph{Object-level usage} Writing extended strategies like \texttt{f(r1[none]\{empty\}, gt-all(match '0.Zero s.t.\ nil))} at the metalevel and executing them with \texttt{metaSrewrite} is possible with what we have introduced so far. However, we want to be able to write them at the object level, like \texttt{f(r1, \skywd{gt-all}(\skywd{match} 0))}, and to use them anywhere a standard strategy can be used, namely, as arguments of the \texttt{srewrite} commands and in strategy definitions within modules. This kind of extensions cannot be directly handled by the Core Maude interpreter, but they can be supported through Full Maude or by custom interactive interfaces~\cite[\S 17]{strategiesClavel,allmaude}. The second option has been chosen.
	
	As stated at the beginning of the section, a parser for the extended strategy language at the object level is required, for which an extensible grammar of the standard strategy language is provided as the \texttt{SLANG-GRAMMAR} in~\cref{fig:extension}. The developer of the extension should complement it with the productions for the new strategy combinators. Moreover, some module-dependent productions available in the standard strategy language (rule labels for their application, sort membership tests, \ldots) can be added with a function provided in the skeleton. Thus, we should extend \texttt{SLANG-GRAMMAR} as we did with \texttt{META-LEVEL}. Using the predefined parsing function \texttt{metaParse}, extended strategy expressions can be parsed and then translated to their representations in the extended metalevel. The skeleton of a \texttt{stratParse} function to convert the terms parsed by this grammar to the \texttt{Strategy} sort is declared and defined for the standard combinators, like the previous \texttt{transform}, so that its definition only has to be completed for the new ones. This closes the procedure depicted in~\cref{fig:extension}. In addition to the grammar of expressions, a limited grammar of strategy modules is already specified to parse those with extended strategies in their definitions, which are translated to the builtin subset in the transformed module. A parametric object-oriented module specifies an interactive interface with an adapted \texttt{srewrite} command, where extended modules can be entered. The interface uses the external objects of Maude~3 for standard input/output communication and the above procedure to parse and execute the strategies.

	For example, the layered normalization strategy \texttt{norm-via-munorm} of~\cref{sec:1stexample} can be specified using a short recursive definition with generic traversals that resumes $\mu$-normalization in all arguments of the term. The translation from equations (and from \texttt{strat} annotations to \texttt{frozen} annotations) in \texttt{LAZY-LIST}, which was automatically done by the previous transformation, has been manually done here.
\begin{lstlisting}
smod LAZY-LIST-STRAT is
  protecting LAZY-LIST-RLS .

  strat norm-via-munorm @ LazyList .
  sd norm-via-munorm := one(all) ! ; gt-all(norm-via-munorm) .
endsm
\end{lstlisting}
The strategy definition is completely generic, although it is parsed in the particular \texttt{LAZY-LIST-RLS} module and translated to the standard subset according to it. In fact, the transformed strategy is essentially the same obtained in~\cref{sec:1stexample}.
\begin{lstlisting}[language={}, escapechar=^]

       ** Strategy language extensions playground **

SLExt> smod LAZY-LIST-STRAT is ^\ldots^ endsm
Module LAZY-LIST-STRAT is now the current module.
SLExt> srew take(3, natsFrom(0)) using norm-via-munorm .
Solution 1: 	0:1:2:nil
No more solutions.
\end{lstlisting}
Another example, where congruence operators are used, is the following module that defines two constants \texttt{a} and \texttt{b}, a binary function \texttt{f}, a rule \texttt{swap} that swaps the entries of \texttt{f}, and another rule \texttt{next} that rewrites \texttt{a} to \texttt{b}.
\begin{lstlisting}
mod FOO is
  sort Foo .
  ops a b : -> Foo [ctor] .
  op  f   : Foo Foo -> Foo [ctor] .

  vars X Y : Foo .
  rl [swap] : f(X, Y) => f(Y, X) .
  rl [next] : a => b .
endm
\end{lstlisting}
The extended strategy \texttt{f(swap, gt-all(next))} can then be executed:
\begin{lstlisting}[language={}]
SLExt> select FOO .
Module FOO is now the current module.
SLExt> srew f(f(a,b), f(a,a)) using f(swap, gt-all(next)) .
Solution 1:     f(f(b, a), f(b, b))
No more solutions.
\end{lstlisting}

	\paragraph{Model checking} In addition to the ability of writing strategy expressions at the object level, another feature of the strategy language we want to preserve is the possibility of model checking systems controlled by strategies. Since extended strategies are finally translated into a strategy expression in the standard strategy language of the original module, model checking with extended strategies is straightforward. In the distribution of the language extension skeleton, a simple Python script makes Maude parse and translate the strategies according to the procedure of~\cref{fig:extension} before passing the problem data to the unified model-checking library \textsf{umaudemc}, where the internal and external model checkers are used to verify LTL, CTL*, and $\mu$-calculus properties (see~\cref{sec:modelchecking}). However, this procedure could have been done entirely in Maude, if we only want to check LTL properties using the builtin model checker.

	\paragraph{Limitations of the approach} This extension procedure can be easily generalized to allow modifications on the subject module where strategies are applied, or to be parametric also on the strategy expression to be evaluated. For example, inline strategy definitions like \texttt{let $\mathit{st}$($t_1$, $\ldots$, $t_n$) := $\beta$ in $\alpha$} can be implemented by pulling the strategy definition in the expression to the target module. However, strategy combinators that are not expressible in the Maude strategy language could not be handled with this approach. Various executable semantics of the strategy language are available to implement other lower-level extensions~\cite{stratweb}.

\section{Multistrategies \refcd{http://maude.ucm.es/strategies/examples/multistrategies.tar.gz}} \label{sec:multistrategies}	

	The strategy-controlled system model proposed in Maude is the combination of a rewrite system and a strategy expression that controls it as a whole. However, many systems are better specified compositionally. Typical examples are object- or agent-oriented systems, in which each object or agent would follow its own strategy. Likewise, describing the interaction of players in games with a single sequential strategy control flow is cumbersome. Hence, we propose the following model transformation to facilitate this specification problem. Instead of a single strategy expression $\alpha$, the system control will be specified by a \emph{multistrategy}: an undetermined number of strategies $\alpha_1, \ldots, \alpha_n$ and a global strategy $\gamma$ that describes how they are combined. Two builtin $\gamma$ are provided: a concurrent one, in which the next strategy to take a step can be any one of them, and a turn-based one, in which strategies are executed in a fixed order.
For example, we could provide each agent of an agent-based system with its own strategy $\alpha_k$ that defines its behavior autonomously. Each strategy can be understood as the program of a concurrent \emph{thread} of execution, which is interrupted and resumed to allow the interleaved interaction of the agents after every rewrite. Alternatively, we could fix an order and make them be executed in turns, as if they were players in a game.
A fundamental question is the amount of atomic work done by a strategy $\alpha_i$ when it is given control, in other words, the granularity of their interleaving in the global execution. A single rule application is a reasonable atomic step, but a few more strategies are executed atomically like \skywd{matchrew}s with a non-trivial pattern and conditions in the conditional operator, since they assume a particular structure or invariant of the term that may not be preserved if another strategy thread modifies the term in the meantime.

	Multistrategies are implemented using strategies at the metalevel and an augmented execution environment. Essentially, to evaluate the strategies $\alpha_1$, \ldots, $\alpha_n$ on the subject term $t$, they are transformed into the term \texttt{\{ $\overline t$ :: < 1 \% $\overline\alpha_1$ > $\cdots$ < $n$ \% $\overline\alpha_n$ >, $\overline M$ \}} that includes the metarepresentation $\overline t$ of the subject term, of the strategies $\overline\alpha_i$, and of the module $\overline M$ where they are evaluated. The evolution of this execution context is defined by some rules, which modify the strategy representations and execute them according to their semantics, governed by the global strategy $\gamma$. The rules that do not alter the subject term (but choose alternatives, expand iterations\ldots) are called \emph{control rules}, and those modifying the term with rules of the underlying system are called \emph{system rules}. These rules and their two categories are directly based on a small-step operational semantics proposed for the strategy language~\cite{fscd}. With \texttt{control(N)} and \texttt{system(N)} being the disjunction of all control and system rules applied to the thread \texttt{N}, global control strategies, like \texttt{turn(N, M)} for executing \texttt{M} strategies in turns starting from the \texttt{N}$^{\mathrm{th}}$ one and \texttt{freec} to execute them concurrently, can be specified as follows:
\begin{lstlisting}
vars N M K : Nat . var T : Term . var  : Strategy .
var Mod : Module .

sd =>>(N) := control(N) * ; system(N) .
sd turns(N, M) := =>>(N) ? turns(s(N) rem M, M) : idle .
sd freec := (matchrew C s.t. { T :: < N by C using =>>(N)) ? freec : idle .
\end{lstlisting}
The \verb|=>>| strategy specifies the atomic step of a strategy thread execution as explained before.\footnote{The concepts of control transition, system transition, and the composed relation \texttt{=>\relax>} are already present in the small-step operational semantics used to define model checking for systems controlled by the Maude strategy language~\cite{fscd}, as both pursue the similar purpose of isolating rule applications.} The definitions of the $\gamma$ strategies \texttt{turns} and \texttt{freec} apply this atomic step \verb|=>>| with indices that are respectively increased cyclically or selected nondeterministically by matching. Note that the \texttt{turns} strategy stops when the current strategy is unable to continue, while \texttt{freec} halts when all threads are stuck. Custom global strategies can be easily defined for other general or specific purposes. For example, the \texttt{freec} strategy can also be bound on the number of steps:
\begin{lstlisting}
sd freec(0) := idle .
sd freec(s(K)) := (matchrew C s.t. { T :: < N := C by C using =>>(N)) ? freec(K) : idle .
\end{lstlisting}
Further details on the transformation are discussed in~\cref{sec:multistrat-impl}, and the complete commented Maude code is available at~\cite{stratweb}.

	Auxiliary operations and an interactive environment have been prepared to easily execute multistrategies at the object level, and to obtain meaningful counterexample traces when model checking these systems. The interactive environment is similar to that used for the language extensions in~\cref{sec:langext}, with a command for rewriting with multistrategies \texttt{srewrite} $t$ \texttt{using} $\alpha_1$\texttt, \ldots\texttt, $\alpha_n$ \texttt{by} $\gamma$, where $\gamma$ can be the words \texttt{turns} or \texttt{concurrent} for the predefined strategies, or \texttt{custom} followed with an arbitrary expression. Another command \texttt{check} $\varphi$ \texttt{from} $t$ \texttt{using} $\alpha_1$\texttt, \ldots\texttt, $\alpha_n$ \texttt{by} $\gamma$ checks the LTL property $\varphi$ on the given multistrategic model. Branching-time properties can also be checked using an external \textsf{umaudemc}-based command line tool.

	Let us illustrate the execution of multistrategies with the simple \texttt{LLIST} example of~\cref{sec:slang}, which specifies a list of letters (\texttt{a}, \texttt{b}, \texttt{c}, \ldots) that can be appended with a \texttt{put} rule, and a strategy \texttt{seq} that does so with a list of them in order.
After loading that module and the interactive interface in \texttt{multistrat-iface.maude}, we can execute multiple \texttt{seq} calls by turns or concurrently:
\begin{lstlisting}[language={}, escapechar=^]
		** Multistrategies playground **

MStrat> select LLIST .
MStrat> srew nil using seq(a b), seq(c d) by turns .
Solution 1: 	a c b d
No more solutions.
MStrat> srew nil using seq(a b), seq(c d) by concurrent .
Solution 1: 	a b c d
^\ldots^
Solution 6: 	c d a b
No more solutions.
\end{lstlisting}

	More interesting examples have been specified using multistrategies~\cite{stratweb}, including the specification of the Lamport's bakery algorithm and the tic-tac-toe game, where relevant properties are model checked using different combinations of process or player strategies. This latter example is studied in the following section.

\subsection{Multistrategies playing games: the tic-tac-toe}

	\emph{Tic-tac-toe} or \emph{noughts and crosses} is a popular game in which two players, circles and crosses, take turns putting their symbols in a 3x3 grid to complete a vertical, horizontal, or diagonal sequence of cells. The first player to achieve the goal is the winner. Tic-tac-toe is a solved game that always ends in a draw if no player makes a mistake. In this section, we will specify a flawless strategy for a player, and use the model checker to prove that it actually is. But before that, the representation of the board and the rules should be specified.
\begin{lstlisting}
fmod TICTACTOE is
  protecting NAT .
  protecting EXT-BOOL .

  sorts Position Player Grid .

  ops O X - : -> Player [ctor] .

  op [_,_,_] : Nat Nat Player -> Grid [ctor] .
  op empty   : -> Grid [ctor] .
  op __      : Grid Grid -> Grid [ctor assoc comm id: empty] .
\end{lstlisting}
The \texttt{Grid} sort's elements are sets of triples that map a coordinate on the game board to the player that occupies that position, \texttt{O} for circles, \texttt{X} for crosses, and \texttt{-} to mean an empty position. In order to decide whether a game is finished, some predicates are defined in order to detect complete rows in every possible direction.
\begin{lstlisting}
  ops hasHRow hasVRow hasDRow : Player Grid -> Bool .
  
  vars I1 I2 I3 J : Nat .  var G : Grid .  var P : Player

  eq hasHRow(P, [I1, J, P] [I2, J, P] [I3, J, P] G) = true .
  eq hasHRow(P, G) = false [owise] .
\end{lstlisting}
As a commutative and associative operator, the grid matches the definition pattern if the entire row \texttt{J} belongs to player \texttt{P} for some \texttt{J}. Predicates for the other directions are defined similarly. Using them, winning is defined by the following disjunction:
\begin{lstlisting}[moredelim={[is][]{\#}{\#}}]
  op hasWon : Player Grid -> Bool .
  eq hasWon(P, G) = hasHRow(P, G) #or-else# hasVRow(P, G)
                    #or-else# hasDRow(P, G) .
endm
\end{lstlisting}
where \texttt{or-else} is a short-circuit version of the logical disjunction provided by the predefined module \texttt{EXT-BOOL}. The game module also defines a constant \texttt{initial} for the initial grid of empty positions \texttt{[$k$, $l$, -]} for all $1 \leq k, l \leq 3$.

	In the system module \texttt{TICTACTOE-RULES}, the player movements are represented by two rules \texttt{putO} and \texttt{putX} that simply place its symbol on an empty position.
\begin{lstlisting}
mod TICTACTOE-RULES is
  protecting TICTACTOE .

  vars I J : Nat .

  rl [putO] : [I, J, -] => [I, J, O] .
  rl [putX] : [I, J, -] => [I, J, X] .
endm
\end{lstlisting}
In these terms, we can define strategies for playing the game in a strategy module \texttt{TICTACTOE-STRAT}. The simplest and most unconscious strategy is the free application of the \texttt{put} rules, but stopping when the game is over. This is what the following \texttt{random} strategies do:
\begin{lstlisting}
smod TICTACTOE-STRAT is
  protecting TICTACTOE-RULES .

  strats randomO randomX @ Grid .
 
  vars G R : Grid . vars I1 I2 I3 I J : Nat . var P : Player .

  sd randomO := (match G s.t. not hasWon(X, G) ; putO) 
                   ? randomO : idle .
  sd randomX := (match G s.t. not hasWon(O, G) ; putX)
                   ? randomX : idle .
\end{lstlisting}
Note that \texttt{random} does not mean that the positions are chosen randomly, but that any of them can be chosen, and so the strategy search commands will explore all possible selections. Moreover, \texttt{randomX} and \texttt{randomO} are not mutually recursive, since each represents the strategy of a different player and turns are handled by the multistrategies framework. These strategies can be improved with certain easy intuitions about more clever movements. For example, if the current player already has two positions in a row and the third one is empty, the symbol should be put there to win immediately. Otherwise, if that situation occurs for the opponent, the active player should occupy the empty position to prevent the other player from winning in its next turn. This is specified by the following \texttt{betterO} strategy for the player \texttt{O} (similarly for the \texttt{X} player).
\begin{lstlisting}
  strats betterO betterX @ Grid .

  sd betterO := (match G s.t. not hasWon(X, G) ;
    ((matchrew G s.t. [I, J, -] R := winningPos(O, G)
        by G using putO[I <- I, J <- J])
    or-else
    (matchrew G s.t. [I, J, -] R := winningPos(X, G)
       by G using putO[I <- I, J <- J])
    or-else
    putO)) ? betterO : idle .
\end{lstlisting}
The \texttt{winningPos($P$, $G$)} function returns a set of sort \texttt{Grid} with all the positions where player $P$ can complete a row. These are calculated equationally, by pattern matching again.
\begin{lstlisting}
  ops winningPos winningHPos winningVPos winningD1Pos
      winningD2Pos : Player Grid -> Grid .

  eq winningPos(P, G) = winningHPos(P, G) winningVPos(P, G)
                    winningD1Pos(P, G) winningD2Pos(P, G) .
  eq winningHPos(P, [I1, J, P] [I2, J, P] [I3, J, -] G) = 
        [I3, J, -] winningHPos(P, G) .
  eq winningHPos(P, G) = empty [owise] .
\end{lstlisting}
Functions to find vertical and diagonal rows are defined similarly. At this point, we can then ask ourselves whether this strategy is \emph{perfect}, i.e.\ whether it always leads to the best possible outcome: not losing the game no matter how the other player behaves. Using the integrated model checker, we can formally verify it. Making \texttt{X} play with \texttt{better} and \texttt{O} play with \texttt{random}, i.e.\ trying all possible moves for \texttt{O}, the property $\ctlAllw \neg \, \mathit{Owins}$ tells us whether \texttt{better} is optimal.
\begin{lstlisting}[language={}, mathescape, escapechar=^, moredelim={[is][\color{blue}]{\#}{\#}}, moredelim={[is][\color{red}]{@}{@}}]
MStrat> select TICTACTOE-CHECK .
Module TICTACTOE-CHECK is now the current module.
MStrat> check [] ~ Owins from initial 
        using betterX, randomO by turns .
@|@ initial
$\color{red}\vee$ #0# does putX
@|@ [1, 1, -] ^\dotfill^ [3, 2, -] [3, 3, X]
$\color{red}\vee$ #1# does putO
@|@ [1, 1, -] ^\dotfill^ [3, 1, -] [3, 2, O] [3, 3, X]
$\color{red}\vee$ #0# does putX
@|@ [1, 1, -] ^\dotfill^ [2, 3, -] [3, 1, X] ^\ldots^ [3, 3, X]
$\color{red}\vee$ #1# does putO
@|@ [1, 1, -] ^\ldots^ [1, 3, -] ^\ldots^ [2, 3, O] ^\dotfill^ [3, 3, X]
$\color{red}\vee$ #0# does putX
@|@ [1, 1, -] ^\ldots^ [1, 3, X] ^\ldots^ [2, 2, -] ^\dotfill^ [3, 3, X]
$\color{red}\vee$ #1# does putO
@|@ [1, 1, -] ^\dotfill^ [2, 1, -] [2, 2, O] ^\dotfill^ [3, 3, X]
$\color{red}\vee$ #0# does putX
@|@ [1, 1, -] [1, 2, -] ^\ldots^ [2, 1, X] ^\dotfill^ [3, 3, X]
$\color{red}\vee$ #1# does putO
^\color{green}\kern1pt X^ [1, 1, -] [1, 2, O] ^\dotfill^ [3, 3, X]
\end{lstlisting}
And it is not, since the counterexample (where we have removed the positions that have not changed) shows an execution in which the circles win even if the crosses play the \texttt{better} strategy. In fact, the only situation where the intelligence of the strategy is actually used is the last move for \texttt{X}, when \texttt{O} has two winning positions in the middle vertical and horizontal row that \texttt{X} cannot block at the same time.

	Hence, \texttt{better} is not a perfect strategy, and further precautions should be taken not to make mistakes. In Table 1 of~\cite{tictactoe}, we can find a script of a perfect strategy for playing tic-tac-toe, which is similar to the algorithm used by the Newell and Simon's tic-tac-toe program in 1972. This script includes various rules to play, which we will call \emph{actions} not to confuse them with rewriting rules, that are not necessarily exclusive. For the strategy to be effective, these actions must be executed in order, applying at each step the first possible one. The first two are those included in \texttt{better}, (1) \action{Win} that completes the row where there are already two positions of the current player, and (2) \action{Block} that prevents the opponent from doing so in the next turn.
\begin{lstlisting}
  strats perfectO perfectX          @ Grid .
  strat  perfect-step      : Player @ Grid .

  sd perfectO := (match G s.t. not hasWon(X, G) ;
       perfect-step(O)) ? perfectO : idle .
  sd perfectX := (match G s.t. not hasWon(O, G) ;
       perfect-step(X)) ? perfectX : idle .

  sd perfect-step(P) :=
    *** Win
    (matchrew G s.t. [I, J, -] R := winningPos(P, G)
       by G using put(P, I, J))
    or-else
    *** Block
    (matchrew G s.t. [I, J, -] R := winningPos(opponent(P), G)
       by G using put(P, I, J))
    or-else
    *** Fork
    (put(P) ; hasFork(P))
    or-else
    *** Blocking an opponent's fork
    (test(put(opponent(P)) ; hasFork(opponent(P))) ; put(P) ;
      *** The opponent cannot fork
      (not(put(opponent(P)) ; hasFork(opponent(P)))
      *** The opponent is forced to block rather that fork
      | (matchrew G s.t. [I, J, -] R := winningPos(P, G)
           by G using not(put(opponent(P), I, J) ;
                          hasFork(opponent(P)))))
    )
    or-else
    *** Center
    put(P, 2, 2)
    or-else
    *** Opposite corner
    ((matchrew [I, I, Q] G s.t. I =/= 2 /\ Q = opponent(P)
        by G using put(P, sd(4, I), sd(4, I)))
    | (matchrew [I, J, Q] G s.t. I =/= 2 /\ J =/= 2 
        /\ Q = opponent(P) by G using put(P, J, I)))
    or-else
    *** Empty corner
    (put(P, 1, 1) | put(P, 3, 3) |
     put(P, 1, 3) | put(P, 3, 1))
    or-else
    *** Empty side
    (match G s.t. P == O ? (putO[I <- 2] | putO[J <- 2])
      : (putX[I <- 2] | putX[J <- 2]))
\end{lstlisting}
The \texttt{or-else} combinator guarantees that actions are applied in order. So as not to define the same strategy twice for each player, as we have done with the previous shorter strategies, the \texttt{perfect-step} strategy takes the player as an argument, and the \texttt{put} strategy and the \texttt{opponent} function have been specified.
\begin{lstlisting}
  strat put : Player Nat Nat @ Grid .
  sd put(X, I, J) := putX[I <- I, J <- J] .
  sd put(O, I, J) := putO[I <- I, J <- J] .
  strat put : Player @ Grid .
  sd put(X) := putX .
  sd put(O) := putO .

  op opponent : Player ~> Player .
  eq opponent(X) = O .
  eq opponent(O) = X .
\end{lstlisting}

After the actions already present in \texttt{better}, the strategy tries (3) \action{Fork} to obtain two winning positions for the next turn, so that winning is guaranteed unless the other player completes a row immediately. Instead of calculating these positions equationally, our strategy puts a symbol randomly and then checks whether there is a fork, with the following strategy:
\begin{lstlisting}
  strat hasFork : Player @ Grid .
  sd hasFork(P) := match G s.t. size(winningPos(P, G)) >= 2 .
\end{lstlisting}
The next actions are (4) \action{Block fork} that prevents the opponent's fork, (5) \action{Center} that occupies the center position, (6) \action{Opposite corner} that fills the diagonally-opposite corner of an opponent's position, (7) \action{Empty corner} that puts the symbol in any corner, and finally (8) \action{Empty side} that uses a side instead. Note that the \emph{Empty side} action is the only remaining possibility when the previous actions have been discarded, so it is equivalent to write simply \texttt{put(P)} instead of the specific strategy we have used in the \texttt{perfect-step} definition.

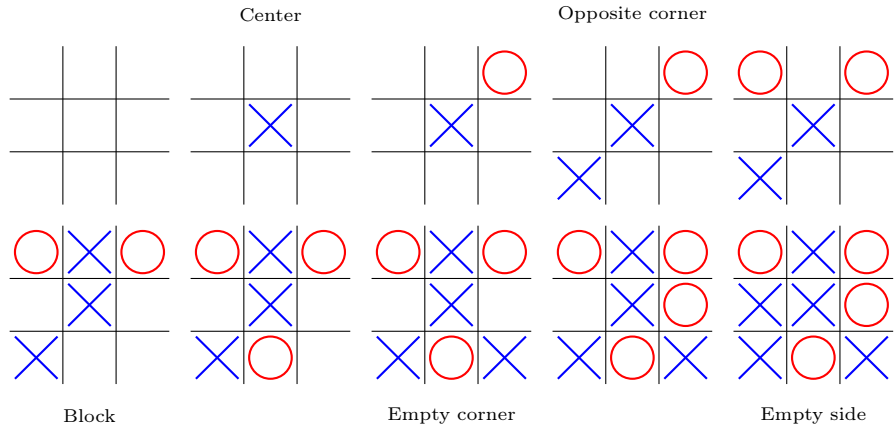
\begin{figure}\centering
\newcommand\dboard[2]{
	\draw (#1, #2 + 2) -- (#1 + 3, #2 + 2);
	\draw (#1, #2 + 1) -- (#1 + 3, #2 + 1);
	\draw (#1 + 1, #2 + 3) -- (#1 + 1, #2);
	\draw (#1 + 2, #2 + 3) -- (#1 + 2, #2);
}
\newcommand\dcross[2]{
	\draw[blue, thick] (#1 - 0.4, #2 - 0.4) -- (#1 + 0.4, #2 + 0.4);
	\draw[blue, thick] (#1 - 0.4, #2 + 0.4) -- (#1 + 0.4, #2 - 0.4);
}
\newcommand\dcircle[2]{\draw[red, thick] (#1, #2) circle (0.4);}
\scriptsize
\begin{tikzpicture}[scale=0.7]
	\foreach \i in {0,...,4} {
		\foreach \j in {0,1} {
			\dboard{3.4 * \i}{3.4 * \j}
		}
	}

\foreach \i in {1,..., 4} {
		\dcross{3.4 * \i + 1.5}{3.4 + 1.5}
	}
	\foreach \i in {0,..., 4} {
		\dcross{3.4 * \i + 1.5}{1.5}
	}

	\node at (3.4 * 1 + 1.5, 7) {Center};

\foreach \i in {2,..., 4} {
		\dcircle{3.4 * \i + 2.5}{3.4 + 2.5}
	}
	\foreach \i in {0,..., 4} {
		\dcircle{3.4 * \i + 2.5}{2.5}
	}

\foreach \i in {3,..., 4} {
		\dcross{3.4 * \i + 0.5}{3.4 + 0.5}
	}
	\foreach \i in {0,..., 4} {
		\dcross{3.4 * \i + 0.5}{0.5}
	}

	\node at (3.4 * 3 + 1.5, 7) {Opposite corner};

\dcircle{3.4 * 4 + 0.5}{3.4 + 2.5}

	\foreach \i in {0,..., 4} {
		\dcircle{3.4 * \i + 0.5}{2.5}
	}

\foreach \i in {0,..., 4} {
		\dcross{3.4 * \i + 1.5}{2.5}
	}

	\node at (3.4 * 0 + 1.5, -.6) {Block};

\foreach \i in {1,..., 4} {
		\dcircle{3.4 * \i + 1.5}{0.5}
	}

\foreach \i in {2,..., 4} {
		\dcross{3.4 * \i + 2.5}{0.5}
	}

	\node at (3.4 * 2 + 1.5, -.6) {Empty corner};

\foreach \i in {3,..., 4} {
		\dcircle{3.4 * \i + 2.5}{1.5}
	}

\dcross{3.4 * 4 + 0.5}{1.5}

	\node at (3.4 * 4 + 1.5, -.6) {Empty side};
\end{tikzpicture}
\caption{Game where \texttt{perfectX} does not win against \texttt{randomO}.} \label{fig:counter}
\end{figure}

	Using the \texttt{check} command and the same property again, we discover that \texttt{perfect} is actually perfect no matter which player starts.
\begin{lstlisting}[language={}]
MStrat> check [] ~ Owins from initial
          using perfectX, randomO by turns .
The property is satisfied.
MStrat> check [] ~ Owins from initial
          using randomO, perfectX by turns .
The property is satisfied.
\end{lstlisting}
However, the strategy does not ensure that \texttt{X} eventually wins. The game may end in a draw, as shown by the following counterexample, which has been drawn in~\cref{fig:counter}.
\begin{lstlisting}[language={}, mathescape, escapechar=^, moredelim={[is][\color{blue}]{\#}{\#}}, moredelim={[is][\color{red}]{@}{@}}]
MStrat> check <> Xwins from initial
          using perfectX, randomO by turns .
@|@ initial
$\color{red}\vee$ #0# does perfect-step
@|@ [1, 1, -] ^\dotfill^ [2, 2, X] ^\dotfill^ [3, 3, -]
$\color{red}\vee$ #1# does putO
@|@ [1, 1, -] ^\dotfill^ [3, 3, O]
$\color{red}\vee$ #0# does perfect-step
@|@ [1, 1, X] ^\dotfill^ [3, 1, -] ^\ldots^ [3, 3, O]
$\color{red}\vee$ #1# does putO
@|@ [1, 1, X] ^\dotfill^ [3, 1, O] [3, 2, -] [3, 3, O]
$\color{red}\vee$ #0# does perfect-step
@|@ [1, 1, X] [1, 2, -] ^\dotfill^ [3, 2, X] [3, 3, O]
$\color{red}\vee$ #1# does putO
@|@ [1, 1, X] [1, 2, O] [1, 3, -] ^\dotfill^ [3, 3, O]
$\color{red}\vee$ #0# does perfect-step
@|@ [1, 1, X] ^\dotfill^ [1, 3, X] ^\dotfill^ [2, 3, -] ^\dotfill^ [3, 3, O]
$\color{red}\vee$ #1# does putO
@|@ [1, 1, X] ^\dotfill^ [2, 1, -] ^\ldots^ [2, 3, O] ^\dotfill^ [3, 3, O]
$\color{red}\vee$ #0# does perfect-step
^\color{green}\kern1pt X^ [1, 1, X] ^\dotfill^ [2, 1, X] ^\dotfill^ [3, 3, O]
\end{lstlisting}
In particular, both players can play the \texttt{perfect} strategy, and then no one wins.
\begin{lstlisting}[language={}]
MStrat> check [] (~ Owins /\ ~ Xwins) from initial 
          using perfectX, perfectO by turns .
The property is satisfied.
\end{lstlisting}

	Finally, we wonder whether the perfect strategy we have adopted from~\cite{tictactoe} is concise or it can be simplified. Repeating the \texttt{check} commands with variations of the strategy, we can see that the last three actions, \action{Opposite corner}, \action{Empty corner}, and \action{Empty side}, can be replaced by the unrestricted \texttt{put(P)}, and so this distinction is superfluous. However, this simplification cannot be extended to the \action{Center} action without losing perfection. Other combinations of rules can be safely removed too.

	Not only the \texttt{check} command is useful in this example, but the \texttt{srewrite} command allows obtaining, for instance, all final configurations of the game using certain strategies.
\begin{lstlisting}[language={}]
MStrat> srew initial using perfectX, randomO by turns .
Solution 1:     [1, 1, -][1, 2, -][1, 3, X]
                [2, 1, -][2, 2, X][2, 3, -]
                [3, 1, X][3, 2, O][3, 3, O]
...
Solution 134:   [1, 1, O][1, 2, X][1, 3, O]
                [2, 1, O][2, 2, X][2, 3, X]
                [3, 1, X][3, 2, O][3, 3, X]
No more solutions
\end{lstlisting}

\subsection{A deeper look into the implementation} \label{sec:multistrat-impl}

Unlike the previous metalevel transformations, multistrategies are not handled by an equational static manipulation of the original module and its strategies. Instead, the term to be rewritten and the multiple strategies that act on it are operated at the metalevel during their execution. The execution context for the multistrategies is not strictly parametric on the subject system, but contains it as data while being rewritten in the system module \texttt{MULTISTRAT}. The already-seen context \texttt{\{ $\overline t$ :: < 1 \% $\overline\alpha_1$ > $\cdots$ < $n$ \% $\overline\alpha_n$ >, $\overline M$ \}} is specified as:
\begin{lstlisting}
sorts   MSContext MSThread MSThreadSet .
subsort MSThread < MSThreadSet .

op <_op none  : -> MSThreadSet [ctor] .
op __    : MSThreadSet MSThreadSet -> MSThreadSet
             [ctor assoc comm id: none] .
op {_::_,_} : Term MSThreadSet Module -> MSContext [ctor] .
\end{lstlisting}
The key fact that lets us follow the execution of the multiple strategies on the subject term is that contexts are univocally associated to terms, and their transitions to their transformations, as we will see.
\begin{lstlisting}
op getTerm : MSContext -> Term .
eq getTerm({ T :: TS, M }) = T .
\end{lstlisting}

	The multiple strategy threads are run on these contexts by several rules that reimplement at some extent the execution of strategies. As said before, some rules only manipulate and decompose strategies, while others may modify the term being rewritten. For example, the following are control transitions for the iteration and disjunction combinators:
\begin{lstlisting}
rl [ms-reduct] : < N rl [ms-reduct] : < N 

crl [ms-choose] : < N if S1 =/= fail /\ S2 =/= fail .
\end{lstlisting}
System transitions are performed by the following rule that executes the strategy \texttt{S}:
\begin{lstlisting}[keywordstyle={[2]{}}]
crl [ms-run] : { T  :: < N => { T' :: < N if S =/= idle
	/\ atomicStrategy(S)
	/\ T' ; Ts := allSuccs(M, T, S) .
\end{lstlisting}
Among the strategies considered \texttt{atomicStrategy} there are not only rule applications, but also \skywd{matchrew} strategies with multiple patterns, as we have said, since they assume a fixed structure of the term along all its execution, which may otherwise be broken by another thread acting on the term. The successors of atomic strategies are calculated using the builtin Maude engine via \texttt{metaSrewrite}. However, to nondeterministically select one of the possible successors, we have to collect all of them in a set with the \texttt{allSuccs} function, and let the rule be instantiated with each by matching the \texttt{E' ; Es} pattern on that set of terms.
Another atomic action is the evaluation of conditions in conditional operators by the following rule:
\begin{lstlisting}
crl [ms-cond] : { T  :: < N => { T' :: < N if T' ; Ts := allSuccs(M, T, S1) .
\end{lstlisting}

Finally, all these rules are gathered in the strategies \texttt{control} and \texttt{system} that lead us to the overview at the beginning of this example.
\begin{lstlisting}[escapechar=^]
strats control system : Nat @ MSContext .

sd system(N) := ms-cond[N <- N] or-else ms-run[N <- N] .
sd control(N) := ms-reduct[N <- N] | ^\ldots^ | ms-def[N <- N] .
\end{lstlisting}
The Maude-based interactive interface and the command-line program to verify branching-time properties are programmed similarly to that for the strategy language extensions seen in~\cref{sec:langext}.

	Unlike the previous examples in~\cref{sec:1stexample,sec:langext}, the reflective implementation of multistrategies just explained operates with the metarepresentation of the strategies at \emph{run time}, instead of producing or \emph{compiling} a new module that is then executed by Maude at the object level. Hence, a sometimes noticeable performance penalty is to be expected when executing and model checking with multistrategies. For example, in the Lamport's bakery algorithm mentioned before and available in the example collection~\cite{stratweb}, generating the entire state space requires 50 seconds with multistrategies but only 200 ms using a less natural alternative strategy that is also included in this specification. However, a more efficient and complex transformation of the first style or a direct implementation could be developed if multistrategies need to scale for more complex applications.

\section{Related work and conclusions} \label{sec:conclusions}

	As we have indicated throughout the paper, the reflective capabilities of Maude have widely been used to build extensions of Maude and frameworks for specific languages and utilities. In addition to Full Maude and the Maude Formal Environment~\cite{mfe}, other relevant examples are Real Time Maude~\cite{realTimeMaude} for specification and verification of real-time systems, and the mobile agents extension Mobile Maude~\cite{mobileMaude}. On the other hand, the strategy language was introduced to control rewriting at the object level without the conceptual difficulties of reflective computations and the intricate shape of metalevel programs. However, some tasks still require resorting to the metalevel, like writing these interactive interfaces or generating strategies depending on the specification or some input data. While Maude has a singular support for reflection and strategies, other strategy languages can also benefit from manipulating and programatically generating strategies as proposed in this paper. The pioneer strategy language ELAN does not originally come with reflective support, but a reflective extension was proposed~\cite{reflectiveElan} where these transformations can be implemented. However, its universal theory does not apparently represent the strategy language combinators themselves, so manipulating strategies could not be straightforward. Partially based on the experience of ELAN, the Porgy graph-rewriting language~\cite{porgyJournal} is reflective since its rewrite rules are graphs themselves that can also be rewritten, but this does not cover its strategy language. Since the Stratego~\cite{stratego} toolset is designed for program transformation, what has been done in this paper could be naturally achieved there. Moreover, some reflective transformations used as examples in this paper would not be necessary in Stratego, since its strategy language is more flexible and supports all the operators we have implemented in~\cref{sec:langext}. In a broader sense, programmable strategies are only general programs whose atomic actions are rule applications, so what is proposed here does not differ much of what could be done in other reflective programming languages. 
	
	 With the examples provided in this paper, we aim to show that manipulating, transforming, and generating strategies is accessible and has useful applications. The reflective representation of the object-level strategy language provides the means to easily do this within Maude, while having strategies executed by the efficient builtin engine. The first example in~\cref{sec:1stexample} shows that strategies can be readily generated to solve specific problems related to program evaluation; the second one in~\cref{sec:langext} allows extending the strategy language with new operators and experiment with them; and the multistrategies of~\cref{sec:multistrategies} can be useful to specify, simulate, and verify systems with distributed control like agent-based or object-oriented systems and games. Another interesting example involving strategy generation is a framework for simulating and verifying membrane systems~\cite{memstratmc}.

\paragraph{Declaration of competing interest}

	The authors declare that they have no known competing financial interests or personal relationships that could have appeared to influence the work reported in this paper.

\paragraph{Acknowledgements}

	Research partially supported by MCI Spanish projects \emph{TRACES} (TIN2015-67522-C3-3-R) and ProCode-UCM (PID2019-108528RB-C22). Rubén Rubio is partially supported by MU grant FPU17/02319.

\bibliographystyle{elsarticle-harv}

\begin{thebibliography}{37}
\expandafter\ifx\csname natexlab\endcsname\relax\def\natexlab#1{#1}\fi
\providecommand{\url}[1]{\texttt{#1}}
\providecommand{\href}[2]{#2}
\providecommand{\path}[1]{#1}
\providecommand{\DOIprefix}{doi:}
\providecommand{\ArXivprefix}{arXiv:}
\providecommand{\URLprefix}{URL: }
\providecommand{\Pubmedprefix}{pmid:}
\providecommand{\doi}[1]{\href{http://dx.doi.org/#1}{\path{#1}}}
\providecommand{\Pubmed}[1]{\href{pmid:#1}{\path{#1}}}
\providecommand{\bibinfo}[2]{#2}
\ifx\xfnm\relax \def\xfnm[#1]{\unskip,\space#1}\fi
\bibitem[{Baader and Nipkow(1998)}]{allthat}
\bibinfo{author}{Baader, F.}, \bibinfo{author}{Nipkow, T.},
  \bibinfo{year}{1998}.
\newblock \bibinfo{title}{Term Rewriting and All That}.
\newblock \bibinfo{publisher}{Cambridge University Press}.
\newblock \DOIprefix\doi{10.1017/CBO9781139172752}.
\bibitem[{Balland et~al.(2007)Balland, Brauner, Kopetz, Moreau and
  Reilles}]{tom}
\bibinfo{author}{Balland, E.}, \bibinfo{author}{Brauner, P.},
  \bibinfo{author}{Kopetz, R.}, \bibinfo{author}{Moreau, P.},
  \bibinfo{author}{Reilles, A.}, \bibinfo{year}{2007}.
\newblock \bibinfo{title}{Tom: {P}iggybacking rewriting on {Java}}, in:
  \bibinfo{editor}{Baader, F.} (Ed.), \bibinfo{booktitle}{Term Rewriting and
  Applications, 18th International Conference, {RTA} 2007, Paris, France, June
  26-28, 2007, Proceedings}, \bibinfo{publisher}{Springer}. pp.
  \bibinfo{pages}{36--47}.
\newblock \DOIprefix\doi{10.1007/978-3-540-73449-9_5}.
\bibitem[{Barendregt(2014)}]{barendregt}
\bibinfo{author}{Barendregt, H.}, \bibinfo{year}{2014}.
\newblock \bibinfo{title}{The Lambda Calculus: Its Syntax and Semantics}.
  volume \bibinfo{volume}{131}.
\newblock \bibinfo{edition}{2} ed., \bibinfo{publisher}{North Holland}.
\bibitem[{Borovansk\'{y} et~al.(2001)Borovansk\'{y}, Kirchner, Kirchner and
  Ringeissen}]{elan}
\bibinfo{author}{Borovansk\'{y}, P.}, \bibinfo{author}{Kirchner, C.},
  \bibinfo{author}{Kirchner, H.}, \bibinfo{author}{Ringeissen, C.},
  \bibinfo{year}{2001}.
\newblock \bibinfo{title}{Rewriting with strategies in {ELAN:} {A} functional
  semantics}.
\newblock \bibinfo{journal}{Int. J. Found. Comput. Sci.} \bibinfo{volume}{12},
  \bibinfo{pages}{69--95}.
\newblock \DOIprefix\doi{10.1142/S0129054101000412}.
\bibitem[{Bourdier et~al.(2009)Bourdier, Cirstea, Dougherty and
  Kirchner}]{extstrat}
\bibinfo{author}{Bourdier, T.}, \bibinfo{author}{Cirstea, H.},
  \bibinfo{author}{Dougherty, D.J.}, \bibinfo{author}{Kirchner, H.},
  \bibinfo{year}{2009}.
\newblock \bibinfo{title}{Extensional and intensional strategies}, in:
  \bibinfo{editor}{Fern\'{a}ndez, M.} (Ed.), \bibinfo{booktitle}{Proceedings
  Ninth International Workshop on Reduction Strategies in Rewriting and
  Programming, {WRS} 2009, Brasilia, Brazil, 28th June 2009}, pp.
  \bibinfo{pages}{1--19}.
\newblock \DOIprefix\doi{10.4204/EPTCS.15.1}.
\bibitem[{Bradfield and Walukiewicz(2018)}]{mucalcmc}
\bibinfo{author}{Bradfield, J.C.}, \bibinfo{author}{Walukiewicz, I.},
  \bibinfo{year}{2018}.
\newblock \bibinfo{title}{The mu-calculus and model checking}, in:
  \bibinfo{editor}{Clarke, E.M.}, \bibinfo{editor}{Henzinger, T.A.},
  \bibinfo{editor}{Veith, H.}, \bibinfo{editor}{Bloem, R.} (Eds.),
  \bibinfo{booktitle}{Handbook of Model Checking}.
  \bibinfo{publisher}{Springer}, pp. \bibinfo{pages}{871--919}.
\newblock \DOIprefix\doi{10.1007/978-3-319-10575-8_26}.
\bibitem[{Bravenboer et~al.(2008)Bravenboer, Kalleberg, Vermaas and
  Visser}]{stratego}
\bibinfo{author}{Bravenboer, M.}, \bibinfo{author}{Kalleberg, K.T.},
  \bibinfo{author}{Vermaas, R.}, \bibinfo{author}{Visser, E.},
  \bibinfo{year}{2008}.
\newblock \bibinfo{title}{Stratego/{XT} 0.17. {A} language and toolset for
  program transformation}.
\newblock \bibinfo{journal}{Sci. Comput. Program.} \bibinfo{volume}{72},
  \bibinfo{pages}{52--70}.
\newblock \DOIprefix\doi{10.1016/j.scico.2007.11.003}.
\bibitem[{Clarke and Emerson(1981)}]{ctl}
\bibinfo{author}{Clarke, E.M.}, \bibinfo{author}{Emerson, E.A.},
  \bibinfo{year}{1981}.
\newblock \bibinfo{title}{Design and synthesis of synchronization skeletons
  using branching-time temporal logic}, in: \bibinfo{editor}{Kozen, D.} (Ed.),
  \bibinfo{booktitle}{Logics of Programs, Workshop, Yorktown Heights, New York,
  USA, May 1981}, \bibinfo{publisher}{Springer}. pp. \bibinfo{pages}{52--71}.
\newblock \DOIprefix\doi{10.1007/BFb0025774}.
\bibitem[{Clavel(2003)}]{strategiesClavel}
\bibinfo{author}{Clavel, M.}, \bibinfo{year}{2003}.
\newblock \bibinfo{title}{Strategies and user interfaces in {Maude} at work},
  in: \bibinfo{editor}{Gramlich, B.}, \bibinfo{editor}{Lucas, S.} (Eds.),
  \bibinfo{booktitle}{Proceedings of the 3rd International Workshop on
  Reduction Strategies in Rewriting and Programming, WRS 2003, Valencia, Spain,
  June 8, 2003}, \bibinfo{publisher}{Elsevier}. pp. \bibinfo{pages}{570--592}.
\newblock \DOIprefix\doi{10.1016/S1571-0661(05)82612-X}.
\bibitem[{Clavel et~al.(2020-10)Clavel, Dur\'{a}n, Eker, Escobar, Lincoln,
  Mart\'{\i}-Oliet, Meseguer, Rubio and Talcott}]{maude}
\bibinfo{author}{Clavel, M.}, \bibinfo{author}{Dur\'{a}n, F.},
  \bibinfo{author}{Eker, S.}, \bibinfo{author}{Escobar, S.},
  \bibinfo{author}{Lincoln, P.}, \bibinfo{author}{Mart\'{\i}-Oliet, N.},
  \bibinfo{author}{Meseguer, J.}, \bibinfo{author}{Rubio, R.},
  \bibinfo{author}{Talcott, C.}, \bibinfo{year}{2020-10}.
\newblock \bibinfo{title}{Maude Manual v3.1}.
\newblock \URLprefix
  \url{http://maude.lcc.uma.es/maude31-manual-html/maude-manual.html}.
\bibitem[{Clavel et~al.(2007a)Clavel, Dur\'{a}n, Eker, Lincoln,
  Mart\'{\i}-Oliet, Meseguer and Talcott}]{allmaude}
\bibinfo{author}{Clavel, M.}, \bibinfo{author}{Dur\'{a}n, F.},
  \bibinfo{author}{Eker, S.}, \bibinfo{author}{Lincoln, P.},
  \bibinfo{author}{Mart\'{\i}-Oliet, N.}, \bibinfo{author}{Meseguer, J.},
  \bibinfo{author}{Talcott, C.L.}, \bibinfo{year}{2007}a.
\newblock \bibinfo{title}{All About {Maude} - {A} High-Performance Logical
  Framework, How to Specify, Program and Verify Systems in Rewriting Logic}.
  volume \bibinfo{volume}{4350} of \textit{\bibinfo{series}{Lecture Notes in
  Computer Science}}.
\newblock \bibinfo{publisher}{Springer}.
\newblock \DOIprefix\doi{10.1007/978-3-540-71999-1}.
\bibitem[{Clavel et~al.(2007b)Clavel, Meseguer and Palomino}]{reflection2007}
\bibinfo{author}{Clavel, M.}, \bibinfo{author}{Meseguer, J.},
  \bibinfo{author}{Palomino, M.}, \bibinfo{year}{2007}b.
\newblock \bibinfo{title}{Reflection in membership equational logic,
  many-sorted equational logic, {Horn} logic with equality, and rewriting
  logic}.
\newblock \bibinfo{journal}{Theor. Comput. Sci.} \bibinfo{volume}{373},
  \bibinfo{pages}{70--91}.
\newblock \DOIprefix\doi{10.1016/j.tcs.2006.12.009}.
\bibitem[{Crowley and Siegler(1993)}]{tictactoe}
\bibinfo{author}{Crowley, K.}, \bibinfo{author}{Siegler, R.S.},
  \bibinfo{year}{1993}.
\newblock \bibinfo{title}{Flexible strategy use in young children's
  tic-tac-toe}.
\newblock \bibinfo{journal}{Cogn. Sci.} \bibinfo{volume}{17},
  \bibinfo{pages}{531--561}.
\newblock \URLprefix \url{10.1016/0364-0213(93)90003-Q}.
\bibitem[{Dur\'{a}n et~al.(2020)Dur\'{a}n, Eker, Escobar, Mart\'{\i}-Oliet,
  Meseguer, Rubio and Talcott}]{maude30}
\bibinfo{author}{Dur\'{a}n, F.}, \bibinfo{author}{Eker, S.},
  \bibinfo{author}{Escobar, S.}, \bibinfo{author}{Mart\'{\i}-Oliet, N.},
  \bibinfo{author}{Meseguer, J.}, \bibinfo{author}{Rubio, R.},
  \bibinfo{author}{Talcott, C.}, \bibinfo{year}{2020}.
\newblock \bibinfo{title}{Programming and symbolic computation in {Maude}}.
\newblock \bibinfo{journal}{J. Log. Algebraic Methods Program.}
  \bibinfo{volume}{110}, \bibinfo{pages}{1--58}.
\newblock \DOIprefix\doi{10.1016/j.jlamp.2019.100497}.
\bibitem[{Dur\'{a}n et~al.(2004)Dur\'{a}n, Escobar and Lucas}]{csrWrla04}
\bibinfo{author}{Dur\'{a}n, F.}, \bibinfo{author}{Escobar, S.},
  \bibinfo{author}{Lucas, S.}, \bibinfo{year}{2004}.
\newblock \bibinfo{title}{New evaluation commands for {M}aude within {F}ull
  {M}aude}, in: \bibinfo{editor}{Mart\'{\i}-Oliet, N.} (Ed.),
  \bibinfo{booktitle}{Proceedings of the Fifth International Workshop on
  Rewriting Logic and its Applications, WRLA 2004, Barcelona, Spain, March
  27-April 4, 2004}, \bibinfo{publisher}{Elsevier}. pp.
  \bibinfo{pages}{263--284}.
\newblock \DOIprefix\doi{10.1016/j.entcs.2004.06.014}.
\bibitem[{Dur\'{a}n et~al.(2007)Dur\'{a}n, Riesco and Verdejo}]{mobileMaude}
\bibinfo{author}{Dur\'{a}n, F.}, \bibinfo{author}{Riesco, A.},
  \bibinfo{author}{Verdejo, A.}, \bibinfo{year}{2007}.
\newblock \bibinfo{title}{A distributed implementation of {M}obile {M}aude},
  in: \bibinfo{editor}{Denker, G.}, \bibinfo{editor}{Talcott, C.} (Eds.),
  \bibinfo{booktitle}{Proceedings of the 6th International Workshop on
  Rewriting Logic and its Applications, WRLA 2006, Vienna, Austria, April 1-2,
  2006}, \bibinfo{publisher}{Elsevier}. pp. \bibinfo{pages}{113--131}.
\newblock \DOIprefix\doi{10.1016/j.entcs.2007.06.011}.
\bibitem[{Dur\'{a}n et~al.(2011)Dur\'{a}n, Rocha and \'{A}lvarez}]{mfe}
\bibinfo{author}{Dur\'{a}n, F.}, \bibinfo{author}{Rocha, C.},
  \bibinfo{author}{\'{A}lvarez, J.M.}, \bibinfo{year}{2011}.
\newblock \bibinfo{title}{Towards a {Maude} {F}ormal {E}nvironment}, in:
  \bibinfo{editor}{Agha, G.}, \bibinfo{editor}{Danvy, O.},
  \bibinfo{editor}{Meseguer, J.} (Eds.), \bibinfo{booktitle}{Formal Modeling:
  Actors, Open Systems, Biological Systems - Essays Dedicated to Carolyn
  Talcott on the Occasion of Her 70th Birthday}, \bibinfo{publisher}{Springer}.
  pp. \bibinfo{pages}{329--351}.
\newblock \DOIprefix\doi{10.1007/978-3-642-24933-4_17}.
\bibitem[{Eker et~al.(2020)Eker, Mart\'{\i}-Oliet, Meseguer, Pita, Rubio and
  Verdejo}]{stratweb}
\bibinfo{author}{Eker, S.}, \bibinfo{author}{Mart\'{\i}-Oliet, N.},
  \bibinfo{author}{Meseguer, J.}, \bibinfo{author}{Pita, I.},
  \bibinfo{author}{Rubio, R.}, \bibinfo{author}{Verdejo, A.},
  \bibinfo{year}{2020}.
\newblock \bibinfo{title}{Strategy language for {Maude}}.
\newblock \URLprefix \url{http://maude.ucm.es/strategies}.
\bibitem[{Eker et~al.(2004)Eker, Meseguer and Sridharanarayanan}]{maudemc}
\bibinfo{author}{Eker, S.}, \bibinfo{author}{Meseguer, J.},
  \bibinfo{author}{Sridharanarayanan, A.}, \bibinfo{year}{2004}.
\newblock \bibinfo{title}{The {Maude} {LTL} model checker}, in:
  \bibinfo{editor}{Gadducci, F.}, \bibinfo{editor}{Montanari, U.} (Eds.),
  \bibinfo{booktitle}{Proceedings of the Fourth International Workshop on
  Rewriting Logic and its Applications, WRLA 2002, Pisa, Italy, September
  19-21, 2002}, \bibinfo{publisher}{Elsevier}. pp. \bibinfo{pages}{162--187}.
\newblock \DOIprefix\doi{10.1016/S1571-0661(05)82534-4}.
\bibitem[{Emerson and Halpern(1986)}]{ctlstar}
\bibinfo{author}{Emerson, E.A.}, \bibinfo{author}{Halpern, J.Y.},
  \bibinfo{year}{1986}.
\newblock \bibinfo{title}{``{S}ometimes'' and ``not never'' revisited: on
  branching versus linear time temporal logic}.
\newblock \bibinfo{journal}{J. {ACM}} \bibinfo{volume}{33},
  \bibinfo{pages}{151--178}.
\newblock \DOIprefix\doi{10.1145/4904.4999}.
\bibitem[{Fernández et~al.(2019)Fernández, Kirchner and
  Pinaud}]{porgyJournal}
\bibinfo{author}{Fernández, M.}, \bibinfo{author}{Kirchner, H.},
  \bibinfo{author}{Pinaud, B.}, \bibinfo{year}{2019}.
\newblock \bibinfo{title}{Strategic port graph rewriting: an interactive
  modelling framework}.
\newblock \bibinfo{journal}{Mathematical Structures in Computer Science}
  \bibinfo{volume}{29}, \bibinfo{pages}{615--662}.
\newblock \DOIprefix\doi{10.1017/S0960129518000270}.
\bibitem[{Kant et~al.(2015)Kant, Laarman, Meijer, van~de Pol, Blom and van
  Dijk}]{LTSmin}
\bibinfo{author}{Kant, G.}, \bibinfo{author}{Laarman, A.},
  \bibinfo{author}{Meijer, J.}, \bibinfo{author}{van~de Pol, J.},
  \bibinfo{author}{Blom, S.}, \bibinfo{author}{van Dijk, T.},
  \bibinfo{year}{2015}.
\newblock \bibinfo{title}{{LTSmin}: High-performance language-independent model
  checking}, in: \bibinfo{editor}{Baier, C.}, \bibinfo{editor}{Tinelli, C.}
  (Eds.), \bibinfo{booktitle}{Tools and Algorithms for the Construction and
  Analysis of Systems, 21st International Conference, TACAS 2015, Held as Part
  of the European Joint Conferences on Theory and Practice of Software, ETAPS
  2015, London, UK, April 11-18, 2015, Proceedings},
  \bibinfo{publisher}{Springer}. pp. \bibinfo{pages}{692--707}.
\newblock \DOIprefix\doi{10.1007/978-3-662-46681-0_61}.
\bibitem[{Kirchner and Moreau(1996)}]{reflectiveElan}
\bibinfo{author}{Kirchner, H.}, \bibinfo{author}{Moreau, P.},
  \bibinfo{year}{1996}.
\newblock \bibinfo{title}{A reflective extension of {ELAN}}, in:
  \bibinfo{editor}{Meseguer, J.} (Ed.), \bibinfo{booktitle}{First International
  Workshop on Rewriting Logic and its Applications, {WRLA} 1996, Asilomar
  Conference Center, Pacific Grove, CA, USA, September 3-6, 1996},
  \bibinfo{publisher}{Elsevier}. pp. \bibinfo{pages}{149--168}.
\newblock \DOIprefix\doi{10.1016/S1571-0661(04)00038-6}.
\bibitem[{Lescanne(1990)}]{lescanneOrme}
\bibinfo{author}{Lescanne, P.}, \bibinfo{year}{1990}.
\newblock \bibinfo{title}{Implementations of completion by transition rules +
  control: {ORME}}, in: \bibinfo{editor}{Kirchner, H.},
  \bibinfo{editor}{Wechler, W.} (Eds.), \bibinfo{booktitle}{Algebraic and Logic
  Programming, Second International Conference, Nancy, France, October 1-3,
  1990, Proceedings}, \bibinfo{publisher}{Springer}. pp.
  \bibinfo{pages}{262--269}.
\newblock \DOIprefix\doi{10.1007/3-540-53162-9_44}.
\bibitem[{Lilis and Savidis(2020)}]{metaprogrammingSurvey}
\bibinfo{author}{Lilis, Y.}, \bibinfo{author}{Savidis, A.},
  \bibinfo{year}{2020}.
\newblock \bibinfo{title}{A survey of metaprogramming languages}.
\newblock \bibinfo{journal}{{ACM} Comput. Surv.} \bibinfo{volume}{52},
  \bibinfo{pages}{113:1--113:39}.
\newblock \DOIprefix\doi{10.1145/3354584}.
\bibitem[{Lucas(2020)}]{contextRew}
\bibinfo{author}{Lucas, S.}, \bibinfo{year}{2020}.
\newblock \bibinfo{title}{Context-sensitive rewriting}.
\newblock \bibinfo{journal}{ACM Comput. Surv.} \bibinfo{volume}{53}.
\newblock \DOIprefix\doi{10.1145/3397677}.
\bibitem[{Lucas(2021)}]{appContextSens}
\bibinfo{author}{Lucas, S.}, \bibinfo{year}{2021}.
\newblock \bibinfo{title}{Applications and extensions of context-sensitive
  rewriting}.
\newblock \bibinfo{journal}{J. Log. Algebraic Methods Program.}
  \bibinfo{volume}{121}.
\newblock \DOIprefix\doi{https://doi.org/10.1016/j.jlamp.2021.100680}.
\bibitem[{Marin and Kutsia(2006)}]{rholog}
\bibinfo{author}{Marin, M.}, \bibinfo{author}{Kutsia, T.},
  \bibinfo{year}{2006}.
\newblock \bibinfo{title}{Foundations of the rule-based system $\rho${L}og}.
\newblock \bibinfo{journal}{J. Appl. Non Class. Logics} \bibinfo{volume}{16},
  \bibinfo{pages}{151--168}.
\newblock \DOIprefix\doi{10.3166/jancl.16.151-168}.
\bibitem[{Meseguer(1992)}]{rewritingLogic}
\bibinfo{author}{Meseguer, J.}, \bibinfo{year}{1992}.
\newblock \bibinfo{title}{Conditional rewriting logic as a unified model of
  concurrency}.
\newblock \bibinfo{journal}{Theor. Comput. Sci.} \bibinfo{volume}{96},
  \bibinfo{pages}{73--155}.
\newblock \DOIprefix\doi{10.1016/0304-3975(92)90182-F}.
\bibitem[{Nieuwenhuis et~al.(2006)Nieuwenhuis, Oliveras and
  Tinelli}]{satstrats}
\bibinfo{author}{Nieuwenhuis, R.}, \bibinfo{author}{Oliveras, A.},
  \bibinfo{author}{Tinelli, C.}, \bibinfo{year}{2006}.
\newblock \bibinfo{title}{Solving {SAT} and {SAT} modulo theories: From an
  abstract {D}avis--{P}utnam--{L}ogemann--{L}oveland procedure to
  {DPLL}(\emph{T})}.
\newblock \bibinfo{journal}{J. {ACM}} \bibinfo{volume}{53},
  \bibinfo{pages}{937--977}.
\newblock \DOIprefix\doi{10.1145/1217856.1217859}.
\bibitem[{\"{O}lveczky(2014)}]{realTimeMaude}
\bibinfo{author}{\"{O}lveczky, P.C.}, \bibinfo{year}{2014}.
\newblock \bibinfo{title}{{Real-Time} {Maude} and its applications}, in:
  \bibinfo{editor}{Escobar, S.} (Ed.), \bibinfo{booktitle}{Rewriting Logic and
  Its Applications - 10th International Workshop, {WRLA} 2014, Held as a
  Satellite Event of ETAPS, Grenoble, France, April 5-6, 2014, Revised Selected
  Papers}, \bibinfo{publisher}{Springer}. pp. \bibinfo{pages}{42--79}.
\newblock \DOIprefix\doi{10.1007/978-3-319-12904-4_3}.
\bibitem[{Pnueli(1977)}]{pneuliLTL}
\bibinfo{author}{Pnueli, A.}, \bibinfo{year}{1977}.
\newblock \bibinfo{title}{The temporal logic of programs}, in:
  \bibinfo{booktitle}{18th Annual Symposium on Foundations of Computer Science,
  Providence, Rhode Island, USA, 31 October - 1 November 1977},
  \bibinfo{publisher}{{IEEE} Computer Society}. pp. \bibinfo{pages}{46--57}.
\newblock \DOIprefix\doi{10.1109/SFCS.1977.32}.
\bibitem[{Rubio(2020)}]{umaudemc}
\bibinfo{author}{Rubio, R.}, \bibinfo{year}{2020}.
\newblock \bibinfo{title}{Unified {Maude} model-checking tool
  (\textsf{umaudemc})}.
\newblock \bibinfo{howpublished}{{FaDoSS}}.
\newblock \URLprefix \url{https://github.com/fadoss/umaudemc}.
\bibitem[{Rubio et~al.(2019a)Rubio, Mart\'{\i}-Oliet, Pita and Verdejo}]{fscd}
\bibinfo{author}{Rubio, R.}, \bibinfo{author}{Mart\'{\i}-Oliet, N.},
  \bibinfo{author}{Pita, I.}, \bibinfo{author}{Verdejo, A.},
  \bibinfo{year}{2019}a.
\newblock \bibinfo{title}{Model checking strategy-controlled rewriting
  systems}, in: \bibinfo{editor}{Geuvers, H.} (Ed.), \bibinfo{booktitle}{4th
  International Conference on Formal Structures for Computation and Deduction,
  {FSCD} 2019, June 24-30, 2019, Dortmund, Germany},
  \bibinfo{publisher}{Schloss Dagstuhl - Leibniz-Zentrum f\"{u}r Informatik}.
  pp. \bibinfo{pages}{34:1--34:18}.
\newblock \DOIprefix\doi{10.4230/LIPIcs.FSCD.2019.31}.
\bibitem[{Rubio et~al.(2019b)Rubio, Mart\'{\i}-Oliet, Pita and Verdejo}]{pssm}
\bibinfo{author}{Rubio, R.}, \bibinfo{author}{Mart\'{\i}-Oliet, N.},
  \bibinfo{author}{Pita, I.}, \bibinfo{author}{Verdejo, A.},
  \bibinfo{year}{2019}b.
\newblock \bibinfo{title}{Parameterized strategies specification in {Maude}},
  in: \bibinfo{editor}{Fiadeiro, J.},
  \bibinfo{editor}{\textcommabelow{T}u\textcommabelow{t}u, I.} (Eds.),
  \bibinfo{booktitle}{Recent Trends in Algebraic Development Techniques},
  \bibinfo{publisher}{Springer}. pp. \bibinfo{pages}{27--44}.
\newblock \DOIprefix\doi{10.1007/978-3-030-23220-7_2}.
\bibitem[{Rubio et~al.(2020a)Rubio, Mart\'{\i}-Oliet, Pita and
  Verdejo}]{memstratmc}
\bibinfo{author}{Rubio, R.}, \bibinfo{author}{Mart\'{\i}-Oliet, N.},
  \bibinfo{author}{Pita, I.}, \bibinfo{author}{Verdejo, A.},
  \bibinfo{year}{2020}a.
\newblock \bibinfo{title}{Simulating and model checking membrane systems using
  strategies in {Maude}}, in: \bibinfo{booktitle}{7th International Workshop on
  Rewriting Techniques for Program Transformation and Evaluation, WPTE 2020,
  Paris, France}, pp. \bibinfo{pages}{1--10}.
\bibitem[{Rubio et~al.(2020b)Rubio, Mart\'{\i}-Oliet, Pita and
  Verdejo}]{btimemc}
\bibinfo{author}{Rubio, R.}, \bibinfo{author}{Mart\'{\i}-Oliet, N.},
  \bibinfo{author}{Pita, I.}, \bibinfo{author}{Verdejo, A.},
  \bibinfo{year}{2020}b.
\newblock \bibinfo{title}{Strategies, model checking and branching-time
  properties in {Maude}}, in: \bibinfo{editor}{Escobar, S.},
  \bibinfo{editor}{Mart\'{\i}-Oliet, N.} (Eds.), \bibinfo{booktitle}{Rewriting
  Logic and Its Applications - 13th International Workshop, {WRLA} 2020,
  Virtual Event, October 20-22, 2020, Revised Selected Papers},
  \bibinfo{publisher}{Springer}. pp. \bibinfo{pages}{156--175}.
\newblock \DOIprefix\doi{10.1007/978-3-030-63595-4_9}.

\end{thebibliography}

\end{document}